\documentclass[usegraphicx, useAMS, usenatbib, letterpaper]{mn2e}
\usepackage{xcolor}
\usepackage[margin=1in]{geometry}
\usepackage{subfigure}
\usepackage{textcomp}
\usepackage{amssymb,array}
\usepackage{graphicx}
\usepackage[fleqn]{amsmath}
\usepackage[T1]{fontenc}
\usepackage[latin9]{inputenc}
\usepackage{float}
\makeatletter
\providecommand{\tabularnewline}{\\}
\floatstyle{ruled}
\newfloat{algorithm}{tbp}{loa}
\providecommand{\algorithmname}{Algorithm}
\floatname{algorithm}{\protect\algorithmname}
\makeatother
\usepackage{xcolor}
\usepackage{ulem}

\begin{document}


\title[Kernel regression time delay estimates]{Kernel regression estimates of time delays between gravitationally lensed fluxes}

\author[S AL Otaibi, P Ti\v{n}o, JC Cuevas-Tello, I Mandel and S Raychaudhury]
{Sultanah AL Otaibi$^{1}$, 
Peter Ti\v{n}o$^{1}$\thanks{P.Tino@cs.bham.ac.uk},
Juan C.~Cuevas-Tello$^2$,  
\newauthor Ilya Mandel$^3$ and
Somak Raychaudhury$^{4,5,3}$\\
$^1$ School of Computer Science, University of Birmingham, Birmingham, B15 2TT, UK\\
$^2$ Engineering Faculty, Autonomous University of San Luis Potosi, M\'{e}xico\\
$^3$ School of Physics and Astronomy, University of Birmingham, Birmingham, B15 2TT, UK\\
$^4$ Inter-University Centre for Astronomy \& Astrophysics, Pun{e} 411007, India\\
$^5$ Department of Physics, Presidency University, Kolkata 700073, India
}

\pagerange{\pageref{firstpage}--\pageref{lastpage}}\pubyear{2015}

\maketitle

\label{firstpage}

\begin{abstract}
Strongly lensed variable quasars can serve as precise cosmological probes, provided that time delays between the image fluxes can be accurately measured.  A number of methods have been proposed to address this problem.  In this paper, we explore in detail a new approach based on kernel regression estimates, which is able to estimate a single time delay given several datasets for the same quasar.  We develop realistic artificial data sets in order to carry out controlled experiments to test of performance of this new approach.  We also test our method on real data from strongly lensed quasar Q0957+561 and compare our estimates against existing results. 
\end{abstract}


\section{Introduction}
Time delays between images of strongly-lensed distant variable sources can serve as a valuable tool for cosmography, providing an alternative to other tools, such as CMB measurements and distance measures based on standard candles \citep[e.g.,][]{Refsdal:1964,Linder:2011:TD,Suyu:2013:TD,Greene:2013:TD,Treu:2013:TD}.  Actively studied strong quasars with time-delay measurements include RXJ1131-1231 \citep[e.g.,][]{Suyu:2013:TD,Tewes:2013:TD2} and B1608+656  \citep[e.g.,][]{Fassnacht:2002,Suyu:2010,Greene:2013:TD}; Q0957+561 \citep[e.g.,][]{Oguri:2007:GLTD, Fadely:2010, Hainline:2012:Q0957}; SDSS J1650+4251 and HE 0435-1223 \citep[e.g.,][]{Kochanek:2006,Vuissoz:2007:COSMOGRAIL,Courbin:2011:COSMOGRAIL}; SDSS J1029+2623 \citep[e.g.,][]{Fohlmeister:2013:TD}; and SDSS J1001+5027 \citep[e.g.,][]{Rathna:2013:COSMOGRAIL}.  These have been used to infer Hubble constant measurements with competitive accuracies. 

However, time delays are difficult to measure because of the unknown intrinsic source variability, the limited observational cadence, and the measurement noise.  A number of methods have been developed to accurately estimate time delays.  These include the dispersion spectra method \citep{Pelt:1998:MDT,Vuissoz:2007:COSMOGRAIL,Courbin:2011:COSMOGRAIL}; the polynomial and curve-fitting methods \citep{Vuissoz:2008:COSMOGRAIL,Eulaers:2013:COSMOGRAIL}; the free-knot spline, variability of regression differences (based on Gaussian process regression), and dispersion minimisation \citep{Tewes:2013:TD}; Gaussian process modelling \citep[e.g.,][]{Hojjati:2013}; and the combined method based on the PRH approach \citep{Hirv:2011:TD}.  However, this remains an active area of research, especially in view of the upcoming surveys such as LSST which will provide unprecedented data sets with strongly lensed distant quasars \citep[e.g.,][]{Treu:2013:TD} \citep[and the recent mock data challenge][]{Dobler:2015,Liao:2015}.

Some of the authors of the present work previously proposed a kernel-based method with variable width (K-V) for time delay estimation \citep{Cuevas-Tello:2006:AA}.  This was combined with an evolutionary algorithm (EA) for parameter optimisation \citep{Cuevas-Tello:2010:PR}. However, the computational time complexity of EA method is $O(n^6)$ \citep{Cuevas-Tello:2007:tesis}. This restriction makes it inadequate for handling  long time series, e.g. \citet{Schild:1997:TD} data\footnote{http://cfa-www.harvard.edu/$\sim$rschild/fulldata2.txt}. This complexity is due to matrix inversion in kernel-based methods for weights estimation. Automatic methods for time delay estimation have been proposed to speed up algorithms in order to deal with long time series, based on Artificial Neural Networks \citep{Gonzalez-Grimaldo:2008:MICAI}; these can be parallelized \citep{Cuevas-Tello:2012:JART}.  Alternatively, a simple hill-climbing optimisation has been proposed \citep{Cuevas:2011:ICAI}.

The main contribution of the present paper is a new probabilistic method that is efficient, robust to observational gaps, capable of directly incorporating measured noise levels reported for individual flux measurement, and able to estimate \textit{a single time delay given several datasets} for the same quasar.  
We also carefully construct synthetic data sets within the framework of multiobjective optimization to reproduce realistic flux variability, observational gaps, and noise levels.  This allows us to test our proposed kernel regression estimate method on synthetic as well as real data, in order to measure the bias and variance of the method.

The paper is organised as follows. Section \S\ref{sec:model} presents the Nadaraya-Watson estimator with known noise levels (henceforth \textit{NWE}) and in \S\ref{sec:full_prob_model} we extend it to a linear noise model with unknown noise (henceforth \textit{NWE++}).   In section  \S\ref{prevwork}, we discuss two previously proposed delay methods, cross correlation and dispersion spectra,  to compare to the new approach. Section \S\ref{sec:data} shows the real datasets studied in this paper, and  presents our procedure for generating synthetic data. The results are given in \S\ref{sec:results}, and we conclude with a summary in \S\ref{sec:conclusion}.

\section{The Model}
\label{sec:model}
We consider a  distant point source (e.g., a quasar) with two strongly lensed images\footnote{generalisation to four images is straightforward}, referred to as $A$ and $B$, and one or more time series of flux measurements, possibly taken by different instruments and/or at different frequencies.  The entire data collection $D = \{ D^1, D^2, ..., D^L\}$ consists of $L$ data sets $D^\ell$, $\ell \in [1,L]$, each corresponding to a sequence of measurements taken with a given instrument and at a given frequency.  Data sets $D^\ell$ consist of flux measurements of both images, $y^\ell_{A}$ and $y^\ell_{B}$, taken at a non-uniform sequence of $N^\ell$ observational times times $t^\ell_1, t^\ell_2, ..., t^\ell_{N^\ell}$.


Formally, each set $D^\ell$ contains $N^\ell$ 3-tuples 

$(t^\ell_k, y^\ell_{A,k}, y^\ell_{B,k})$, $k=1,2,...,N^\ell$,
\begin{small}
\[
D^\ell = \{ (t^\ell_1, y^\ell_{A,1}, y^\ell_{B,1}), 
(t^\ell_2, y^\ell_{A,2}, y^\ell_{B,2}), ...,
(t^\ell_{N^\ell}, y^\ell_{A,{N^\ell}}, y^\ell_{B,{N^\ell}})  \},
\]
\end{small}
where $y^\ell_{A,k}$ and $y^\ell_{B,k}$ denote the observed fluxes
of image A and B, respectively, in $D^\ell$ at time $t^\ell_k$. We also assume that the standard errors $\sigma^\ell_{A,k}$ and $\sigma^\ell_{B,k}$ are known for each observation $y^\ell_{A,k}$ and $y^\ell_{B,k}$, respectively.

The fluxes corresponding to the two images A and B are collected in sets 
\[
D_A^\ell = \{ (t^\ell_1, y^\ell_{A,1}), 
(t^\ell_2, y^\ell_{A,2}), ...,
(t^\ell_{N^\ell}, y^\ell_{A,{N^\ell}})  \}
\]
and
\[
D_B^\ell = \{ (t^\ell_1, y^\ell_{B,1}), 
(t^\ell_2, y^\ell_{B,2}), ...,
(t^\ell_{N^\ell}, y^\ell_{B,{N^\ell}})  \}.
\]

For observations at frequencies above a few tens of MHz, dispersion yields sub-hour arrival time differences, and is not significant relative to typical time-delay measurement accuracy.  We therefore assume that the time delay between gravitationally lensed fluxes does not depend on the wavelength at which the observations are taken. We also assume stationarity of the lensing object (e.g., a galaxy) in the sense that the delay does not change in time; in particular, we ignore micro-lensing contributions. 

\subsection{Nadaraya-Watson Estimator with Known Noise Levels (NWE)}

Given a delay $\Delta$, we seek to find a probabilistic model $p(D | \Delta)$ that explains\footnote{
We slightly abuse mathematical notation as we are actually building conditional models of flux values, given the observation times.}
 $D$. Assuming independence of the observation sets $D^\ell$, we obtain
\[
p(D | \Delta) = \prod_{\ell=1}^L p(D^\ell | \Delta).
\]
Assuming independent observations at distinct measurement times, we get
\[
p(D^\ell | \Delta) = \prod_{k=1}^{N^\ell} 
p(y^\ell_{A,k}, y^\ell_{B,k} | t^\ell_k, \Delta)
\]
and further assumption of independence of measurement noise in images A and B leads to
\[
p(y^\ell_{A,k}, y^\ell_{B,k} | t^\ell_k, \Delta)
=
p_A(y^\ell_{A,k} | t^\ell_k, \Delta) \
p_B(y^\ell_{B,k} | t^\ell_k, \Delta).
\]

\subsubsection{Modelling the source using image A}
\label{sec:im_A}

It is typically assumed that the measurement uncertainties on fluxes $D_A^\ell$ and $D_B^\ell$ are normally distributed, with zero mean Gaussian noise of known standard deviation $\sigma^\ell_{A,k}$ and $\sigma^\ell_{B,k}$ associated with noisy observations $y^\ell_{A,k}$ and $y^\ell_{B,k}$, respectively. We model the mean of image A using Nadaraya-Watson kernel regression \citep{Nadaraya:1964}, \citep{Watson:1964},
\begin{equation}
f^{\ell}_A(t) = \sum_{k=1}^{N^\ell} 
y^\ell_{A,k} \ 
\frac{K(t, t^\ell_k; h^\ell)}
{\sum_{j=1}^{N^\ell} K(t, t^\ell_j; h^\ell)},
\label{eq:f_A}
\end{equation}
where $f^{\ell}(t)$ is the predicted flux at time $t$ and $K(t, t^\ell_j; h^\ell)$ is a kernel positioned at $t^\ell_j$ with bandwidth parameter $h^\ell$. We use the Gaussian kernel
\[
K(t, t_k; h) = \exp\left\{ - \frac{(t-t_k)^2}{\kappa^2(t_k)} \right\},
\]
where the kernel scale $\kappa(t_k)$ at position $t_k$ is defined as the distance spanned by the $h$ neighbours (to the left and to the right) of $t_k$, i.e.
$\kappa(t_k) = t_{k+h} - t_{k-h}$.
{This approach to modelling the noise should work when the autocorrelation length of the observed flux is much longer than any gaps in the data during which the flux is modelled via the Nadaraya-Watson kernel regression estimator.  If the autocorrelation length of the observed flux, which can be estimated from a time interval when the observations are relatively closely spaced, is comparable to or larger than a data gap, this approach (or any other approach that does not incorporate a physically accurate flux model) cannot be trusted.}

To respect the nature of gravitationally lensed data, we impose that the mean model for image B follows exactly that for image A, up to scaling by a constant\footnote{assumed known, or easily estimated in a preprocessing stage using the means of the fluxes in $D_A^\ell$ and $D_B^\ell$} 
$M>0$ and time shift by $\Delta$:
\[
f^{\ell}_B(t; \Delta) = M \ f^{\ell}_A(t-\Delta).
\] 

Since the shift $\Delta$ plays no role in modelling image A, we write
\small
\begin{equation}
p(y^\ell_{A,k}, y^\ell_{B,k} | t^\ell_k, \Delta)
=
p_A(y^\ell_{A,k} | t^\ell_k) \
p_B(y^\ell_{B,k} | t^\ell_k, \Delta),
\end{equation}
\normalsize 
where 
\small
\begin{equation}
p_A(y^\ell_{A,k} | t^\ell_k) = \frac{1}{\sqrt{2 \pi} \ \sigma^\ell_{A,k}} \ 
\exp\left\{ - \frac{1}{2} \frac{(y^\ell_{A,k} - f^{\ell}_A(t^\ell_k))^2}{(\sigma^\ell_{A,k})^2} \right\}
\label{eq:P_A}
\end{equation}
\normalsize 
and
\small
\begin{equation}
\begin{split}
p_B(y^\ell_{B,k} | t^\ell_k, \Delta) 
&= \frac{1}{\sqrt{2 \pi} \ \sigma^\ell_{B,k}} \ \\
&  \quad \cdot\exp\left\{ - \frac{1}{2} \frac{(y^\ell_{B,k} - M f^{\ell}_A(t^\ell_k - \Delta))^2}{(\sigma^\ell_{B,k})^2} \right\}.
\label{eq:P_B}
\end{split}
\end{equation}
\normalsize 
Note that given $\Delta$, the only free parameter of 
$p(y^\ell_{A,k}, y^\ell_{B,k} | t^\ell_k, \Delta)$ is the kernel width parameter $h^\ell$
in the formulation of the mean model (\ref{eq:f_A}).

Ignoring constant terms and scaling, the negative log likelihood, $- \log p(D^\ell|\Delta)$, forms the approximation error for the set $D^\ell$,
\small
\begin{equation}
\begin{split}
E^\ell_A(h^\ell; \Delta)
& = \sum_{k=1}^{N^\ell}
\left\{ 
\frac{(y^\ell_{A,k} - f^{\ell}_A(t^\ell_k))^2}{(\sigma^\ell_{A,k})^2} \right. \\
&  \quad+\left.\frac{(y^\ell_{B,k} - M f^{\ell}_A(t^\ell_k - \Delta))^2}{(\sigma^\ell_{B,k})^2}
\right\}.
\label{eq:E_A}
\end{split}
\end{equation}
\normalsize 

Writing down the negative log likelihood for the whole data,
$- \log p(D|\Delta)$, and ignoring scaling and constant terms leads to
the total approximation error
\small
\[
E_A(\textbf{h}; \Delta) = 
\sum_{\ell=1}^{L} E^\ell_A(h^\ell; \Delta),
\]
\normalsize 
where $\textbf{h} = (h^1, h^2, ..., h^L)$ is a vector that collects kernel width parameters for all datasets $D^1, D^2, ..., D^L$ in $D$.

\subsubsection{Modelling the source using image B}

One can, of course, start by building a mean flux model $f^{\ell}_B(t)$ for image B via Nadaraya-Watson kernel regression, 
\small
\begin{equation}
f^{\ell}_B(t) = \sum_{k=1}^{N^\ell} 
y^\ell_{B,k} \ 
\frac{K(t, t^\ell_k; h^\ell)}
{\sum_{j=1}^{N^\ell} K(t, t^\ell_j; h^\ell)},
\label{eq:f_B}
\end{equation}
\normalsize 
imposing that the mean model of image A is
\small
\[
f^{\ell}_A(t; \Delta) = \frac{1}{M} \ f^\ell_B(t+\Delta).
\]
\normalsize 

Crucially, since both images A and B come from the same source, we require that the kernel width $h^\ell$ for the mean models 
$f^{\ell}_A(t)$ and $f^{\ell}_B(t)$ (and hence for $f^{\ell}_A(t; \Delta)$ and $f^{\ell}_B(t; \Delta)$ as well) be the same for the whole dataset
$D^\ell$.

Using the same reasoning as in section \S\ref{sec:im_A}, we obtain an approximation error for the set $D^\ell$:
\footnotesize
\[
E^\ell_B(h^\ell; \Delta) = 
\sum_{k=1}^{N^\ell}
\left\{ 
\frac{(y^\ell_{A,k} - \frac{1}{M} f^{\ell}_B(t^\ell_k + \Delta))^2}{(\sigma^\ell_{A,k})^2} \\
 + \frac{(y^\ell_{B,k} - f^{\ell}_B(t^\ell_k))^2}{(\sigma^\ell_{B,k})^2} 
\right\}
\]
\normalsize 
leading to the total approximation error
\small
\[
E_B(\textbf{h}; \Delta) = 
\sum_{\ell=1}^{L} E^\ell_B(h^\ell; \Delta).
\]
\normalsize 

\subsubsection{Estimating the Unique Time Delay across $D$}
\label{ref:est_Delta}
Since there is no a-priori reason to prefer one image over the other, we aim to find the unique delay $\Delta$ that minimises both the errors $E_A(\textbf{h}; \Delta)$ and $E_B(\textbf{h}; \Delta)$ with the same `level of importance'. In other words, 
we are looking for $\Delta$ and the set of kernel width parameters $\textbf{h} = (h^1, h^2, ..., h^L)$, one for each dataset $D^\ell$ in $D$, that minimise the error
\small
\[
E(\textbf{h}; \Delta) = E_A(\textbf{h}; \Delta) + E_B(\textbf{h}; \Delta).
\]
\normalsize 

Note that the imposition that there is a unique delay $\Delta$ for the whole data $D$ and that the kernel widths are the same throughout each set $D^\ell$ for all the corresponding mean models $f^{\ell}_A(t)$, $f^{\ell}_B(t)$, $f^{\ell}_A(t; \Delta)$ and $f^{\ell}_B(t; \Delta)$, not only makes sense from the point of view of underlying physics, but is also a stabilising factor in the analysis and modelling of $D$. 

The structure of our problem enables us to use an efficient and practical approach to finding the optimal time delay $\Delta_*$.
The error $E(\textbf{h}; \Delta)$ to be minimised can be rewritten as
\small
\begin{equation}
\label{eq:unique-delay} 
E(\textbf{h}; \Delta) = 
\sum_{\ell=1}^{L} E^\ell(h^\ell; \Delta),
\end{equation}
\normalsize
where
\small
\[
E^\ell(h^\ell; \Delta) =
E^\ell_A(h^\ell; \Delta) + E^\ell_B(h^\ell; \Delta).
\]
\normalsize

For every test value $\Delta$ we can separately optimise
$E^\ell(h^\ell; \Delta)$ for $h^\ell$ within each set $D^\ell$. Note that this boils down into a set of $L$ one-dimensional optimisations of bandwidths $h^1, h^2,...,h^L$. In addition, because of the nature of the mean models, the errors $E^\ell(h^\ell; \Delta)$ will behave `reasonably' with changes in $h^\ell$, i.e. the changes will be smooth and we can expect {a} roughly unimodal shape of cross-validated $E^\ell(h^\ell; \Delta)$. That enables us to use further speed-up tricks (such as halving) in the 1-dimensional optimisations.
The estimated time delay is the one with the minimal overall $E(\textbf{h}; \Delta)$ for the (cross-validation) optimised kernel width parameters $\textbf{h}$.

\section{Nadaraya-Watson Estimator with Linear Noise Model (NWE++)}
\label{sec:full_prob_model}
In section \S\ref{sec:model} only the mean fluxes were modelled, the standard errors on observations were assumed known. Our approach can be extended to full probabilistic modelling by assuming a model for the relationship between the noise level and the observed fluxes.  Here, we consider a simple model in which the standard error on the measured flux depends linearly on the observed flux value $y$, i.e., $\sigma(y) = \nu \cdot y$, where the proportionality constant $\nu$ depends on the wavelength at which the flux is measured (e.g., $\nu$ could be 1\% and 0.1\% for radio and optical data, respectively).  Assuming that the mean models for dataset $D^\ell$ are fitted reasonably well, so that $y^\ell_{I,k} \approx f^\ell_I(t^\ell_k)$, $I \in \{ A,B \}$, then {to lowest order} 
$\sigma(y^\ell_{I,k}) \approx \nu^\ell \cdot f^\ell_I(t^\ell_k)$.  

Most of the material developed in sections \ref{sec:model} will stay unchanged, modifications are required only in the formulation of the noise models (\ref{eq:P_A}) and (\ref{eq:P_B}):
\small
\begin{equation}
p_A(y^\ell_{A,k} | t^\ell_k) = \frac{1}{\nu^\ell \ \sqrt{2 \pi} \ f^\ell_A(t^\ell_k)} \ 
\exp\left\{
\frac{-1}{2(\nu^\ell)^2} \left[ \frac{y^\ell_{A,k}}{f^{\ell}_A(t^\ell_k)} - 1 \right]^2 
\right\}
\label{eq:P_A_f}
\end{equation}
\normalsize
and
\small
\begin{equation}
\begin{split}
p_B(y^\ell_{B,k} | t^\ell_k, \Delta) 
&= 
\frac{1}{M \nu^\ell \ \sqrt{2 \pi} \ f^\ell_A(t^\ell_k-\Delta)} \ \\
&  \quad \cdot \exp\left\{ 
\frac{-1}{2(\nu^\ell)^2} \left[ \frac{y^\ell_{B,k}}{M f^{\ell}_A(t^\ell_k-\Delta)} - 1 \right]^2 
\right\}.
\label{eq:P_B_f}
\end{split}
\end{equation}
\normalsize

This time, however, we can write a full probabilistic model for any time point $t$ and evaluate the likelihood within our model given any observation pair
$(y^\ell_A(t),y^\ell_B(t))$ that could have been measured at time $t$:
\small
\begin{equation}
p_A(y^\ell_A(t)) = \frac{1}{\nu^\ell \sqrt{2 \pi} \ f^\ell_A(t)} \ 
\exp\left\{
\frac{-1}{(\nu^\ell)^2} \left[ \frac{y^\ell_A(t)}{f^{\ell}_A(t)} - 1 \right]^2 
\right\}
\label{eq:P_A_f_t}
\end{equation}
\normalsize
and
\small
\begin{equation}
\begin{split}
p_B(y^\ell_B(t) | \Delta) 
&= 
\frac{1}{M \nu^\ell \ \sqrt{2 \pi} \ f^\ell_A(t-\Delta)} \\
&
\cdot \exp\left\{ 
\frac{-1}{(\nu^\ell)^2} \left[ \frac{y^\ell_B(t)}{M f^{\ell}_A(t-\Delta)} - 1 \right]^2 
\right\}.
\label{eq:P_B_f_t}
\end{split}
\end{equation}
\normalsize

The approximation error $E^\ell_A(h^\ell; \Delta)$ to be minimised by the choice of kernel width $h^\ell$ now reads:
\scriptsize
\[
E^\ell_A(h^\ell; \Delta) 
=
\frac{1}{(\nu^\ell)^2}
\sum_{k=1}^{N^\ell}
\left\{ 
\left[ \frac{y^\ell_{A,k}}{f^{\ell}_A(t^\ell_k)} - 1 \right]^2  
 + \left[ \frac{y^\ell_{B,k}}{M f^{\ell}_A(t^\ell_k-\Delta)} - 1 \right]^2 
\right\}.
\]
\normalsize

Following analogous arguments for the case of modelling the source using image B, we have
\small
\begin{equation}
\begin{split}
p_A(y^\ell_A(t) | \Delta) 
&= 
\frac{M}{\nu^\ell \sqrt{2 \pi} \ f^\ell_B(t+\Delta)} \\ 
& \cdot \exp\left\{ 
\frac{-1}{(\nu^\ell)^2} \left[ \frac{M \ y^\ell_A(t)}{f^{\ell}_B(t+\Delta)} - 1 \right]^2 
\right\}
\label{eq:P_A_f_t_B}
\end{split}
\end{equation}
\normalsize
and
\small
\[
p_B(y^\ell_B(t)) = \frac{1}{\nu^\ell \sqrt{2 \pi} \ f^\ell_B(t)} \ 
\exp\left\{
\frac{-1}{(\nu^\ell)^2} \left[ \frac{y^\ell_B(t)}{f^{\ell}_B(t)} - 1 \right]^2 
\right\},
\]
\normalsize
which leads to the approximation error
\small
\[
E^\ell_B(h^\ell; \Delta) 
=
\frac{1}{(\nu^\ell)^2}
\sum_{k=1}^{N^\ell}
\left\{ 
\left[ \frac{M \ y^\ell_{A,k}}{f^{\ell}_B(t^\ell_k + \Delta)} - 1 \right]^2 
+
\left[ \frac{y^\ell_{B,k}}{f^{\ell}_B(t^\ell_k)} - 1 \right]^2 
\right\}.
\]
\normalsize

Again, the final cost to be minimised is
\small
\begin{equation}
\label{eq:final-cost}
E(\textbf{h}; \Delta) = 
\sum_{\ell=1}^{L} E^\ell(h^\ell; \Delta),
\end{equation}
\normalsize
where
\small
\[
E^\ell(h^\ell; \Delta) =
E^\ell_A(h^\ell; \Delta) + E^\ell_B(h^\ell; \Delta).
\]
\normalsize

\section{Previous Work}\label{prevwork}
\subsection{Cross Correlation}\label{cross-correlation-section}
There are two versions of the methods based on cross correlation: the Discrete Correlation Function (DCF) \citep{Edelson:1988:DCF} and its variant, the Locally Normalised Discrete Correlation Function (LNDCF) \citep{Lehar:1992:TRT}. Both calculate correlations directly on discrete pairs of light curves. These methods avoid interpolation in the observational gaps. They are also the simplest and quickest time delay estimation methods.

First, time differences (lags), $\Delta t_{ij} = t_j - t_i$, between all pairs of observations are binned into discrete bins. Given a bin size $\Delta \tau$, the bin centred at lag $\tau$ is the time interval $I_{\tau}=[\tau - \Delta \tau / 2,  \tau+\Delta\tau/2]$. The DCF at lag $\tau$ is given by
\small
\begin{equation}
  \label{dcf}
   DCF(\tau) = \frac{1}{P(\tau)} \sum_{i,j}^{t_i,t_j \in I_{\tau}} \frac{(y_A(t_i)- \bar{a})(y_B(t_j)- \bar{b}) }{\sqrt{(\sigma_a^2-\sigma_A^2(t_i))(\sigma_b^2-\sigma_B^2(t_j))} },
\end{equation}
\normalsize
\noindent where $P(\tau)$ is the number of observational pairs in the bin centred at $\tau$, $\bar{a}$ and $\bar{b}$ are the means of the observed data, $y_A(t_i)$ and $y_B(t_j)$, and their variances are $\sigma_a^2$ and $\sigma_b^2$, respectively.

Likewise, 
\small
\begin{equation}  
\begin{split}
\label{lndcf}
   LNDCF(\tau) 
   &= \frac{1}{P(\tau)} \\
   & \cdot\sum_{i,j}^{t_i,t_j \in I_{\tau}} 
   \frac{(y_A(t_i)- \bar a(\tau))(y_B(t_j)-  \bar b(\tau)) }
   {\sqrt{(\sigma^2_a(\tau)-\sigma_A^2(t_i))
   (\sigma^2_b(\tau)-\sigma_B^2(t_j))} },
\end{split}
\end{equation}
\normalsize

\noindent  where $\bar a(\tau)$, $\bar b(\tau)$, $\sigma^2_a(\tau)$ and $\sigma^2_b(\tau)$ are the lag means and variances in the bin centred  at $\tau$. 

The time delay $\Delta$ is found when $DCF(\tau)$ and $LNDCF(\tau)$, given by equations (\ref{dcf}) and (\ref{lndcf}), are greatest; i.e., at the best correlation \citep{Edelson:1988:DCF,Lehar:1992:TRT}.

\subsection{Dispersion Spectra}\label{dispersion-section}

The Dispersion Spectra method \citep{Pelt:1996:TLC,Pelt:1998:MDT} 
measures the dispersion of time series of two light curves $y_A(t_i) $ and $y_B(t_j) $ 
by combining them (given a trial time delay $\Delta$ and ratio $M$) into a single signal, 
$y(t_k)$, $k=1,2,..., 2N$. In other words, given the delay $\Delta$, the observed values of signal $A$,
$\{y_A(t_i)\}_{i=1}^N $, and (delayed and rescaled) signal $B$, $ \{\tilde y_B(t_i)\}_{i=1}^N$,
where $\tilde y_B(t) = M y_B(t-\Delta)$, are joined together and re-ordered in time, forming a joint signal  $\{y(t_k)\}_{k=1}^{2N} $ of length $2N$.
We employ two versions of this method \citep{Pelt:1998:MDT}:
\small
\begin{equation}
  \label{D1-equation}
   DS_{1}^{2}(\Delta) = \!\min_{M} \frac{\sum_{a=1}^{2N-1}w_{a} \left( y(t_{a+1})-y(t_{a}) \right) ^{2}}{2\sum_{a=1}^{2n-1}w_{a}}
\end{equation}
\normalsize
\noindent and
\begin{scriptsize}
\begin{equation}
  \label{D4-equation}
   DS_{2,4}^{2}(\Delta) = \!\min_{M} \frac{\sum_{a=1}^{2N-1}\sum_{c=a+1}^{2N}H_{a,c} W_{a,c} G_{a,c} \left( y(t_{a})-y(t_{c}) \right) ^{2}}{2\sum_{a=1}^{2n-1}\sum_{c=a+1}^{2n}H_{a,c} W_{a,c} G_{a,c}},
\end{equation}
\end{scriptsize}
where
\begin{small}
\begin{equation}
  \label{W-equation}
   w_{a}=\frac{1}{\sigma^2(t_{a+1})+\sigma^2(t_{a})}, \\  W_{a,c}=\frac{1}{\sigma^2(t_{a})+\sigma^2(t_{c})} 
\end{equation}
\end{small}

\noindent are the statistical weights taking in account the measurement errors, where $G_{a,c}=1$ only when $y(t_{a})$ and $y(t_{c})$ are from different images, and $G_{a,c}=0$ otherwise, and
\small
\begin{equation}
  \label{S-equation}
   H_{a,c}= \left\{ \begin{array}{ll}
                                1-\frac{\vert t_{a} - t_{c} \vert}{\delta}, & \hbox{if}\ \ \vert t_{a} - t_{c} \vert \leq \delta\\
                                0, & \hbox{otherwise}.
                          \end{array}
                  \right.
\end{equation}
\normalsize

Compared with $DS_{1}^{2}$, the $DS_{2,4}^{2}$ method has an additional parameter, the {\it decorrelation length} $\delta$, which signifies the maximum distance between observations that we are willing to consider when calculating the correlations \citep{Pelt:1996:TLC}.

\noindent The estimated time delay $\Delta$ is found by minimising $DS^{2}$
over a range of time delay trials $\Delta$, as above.

\section{Data}
\label{sec:data}
\begin{figure*}
 \centering
  \subfigure[$D^1$]{\includegraphics[width=3.1in]{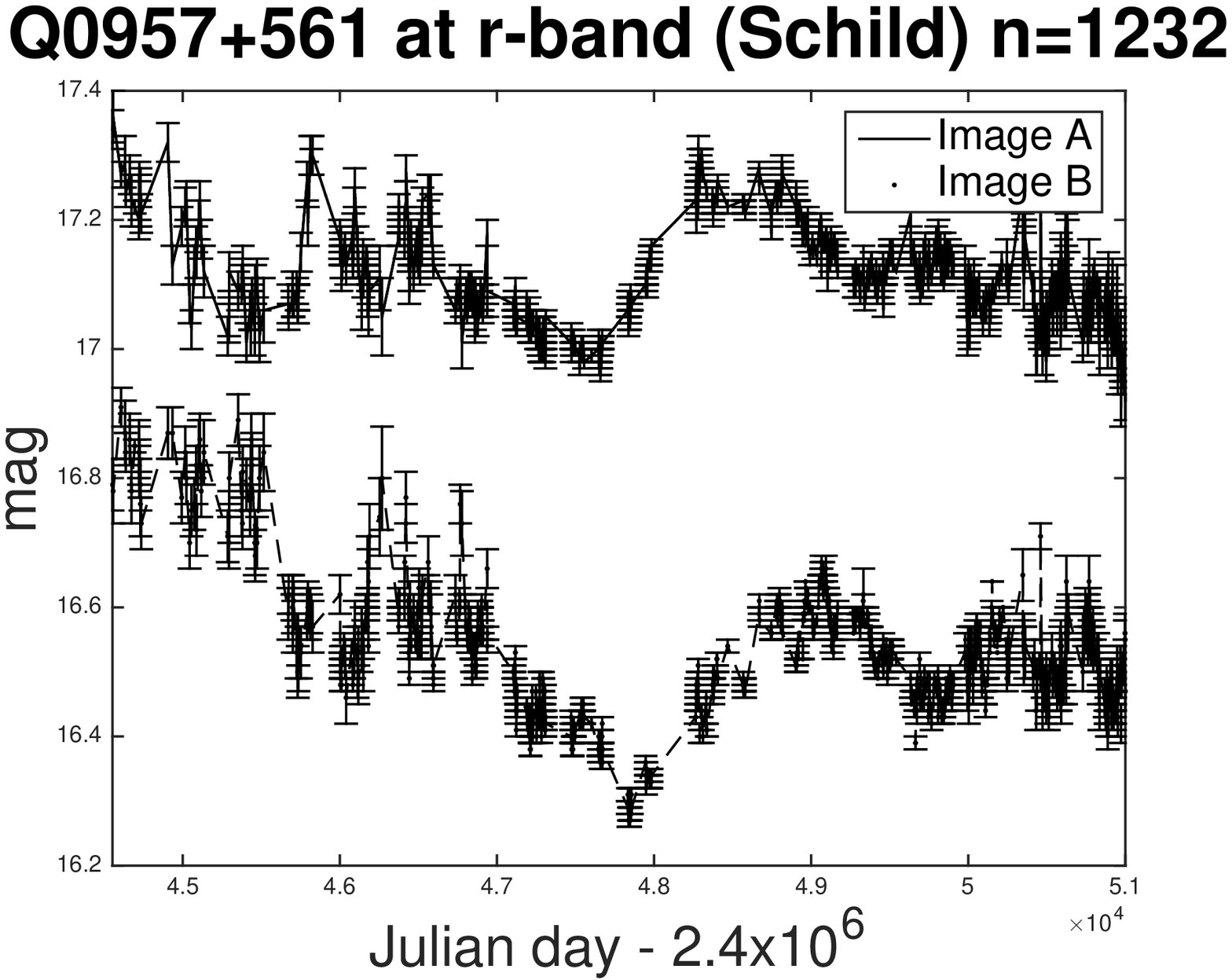}}
  \subfigure[$D^2$]{\includegraphics[width=3.1in]{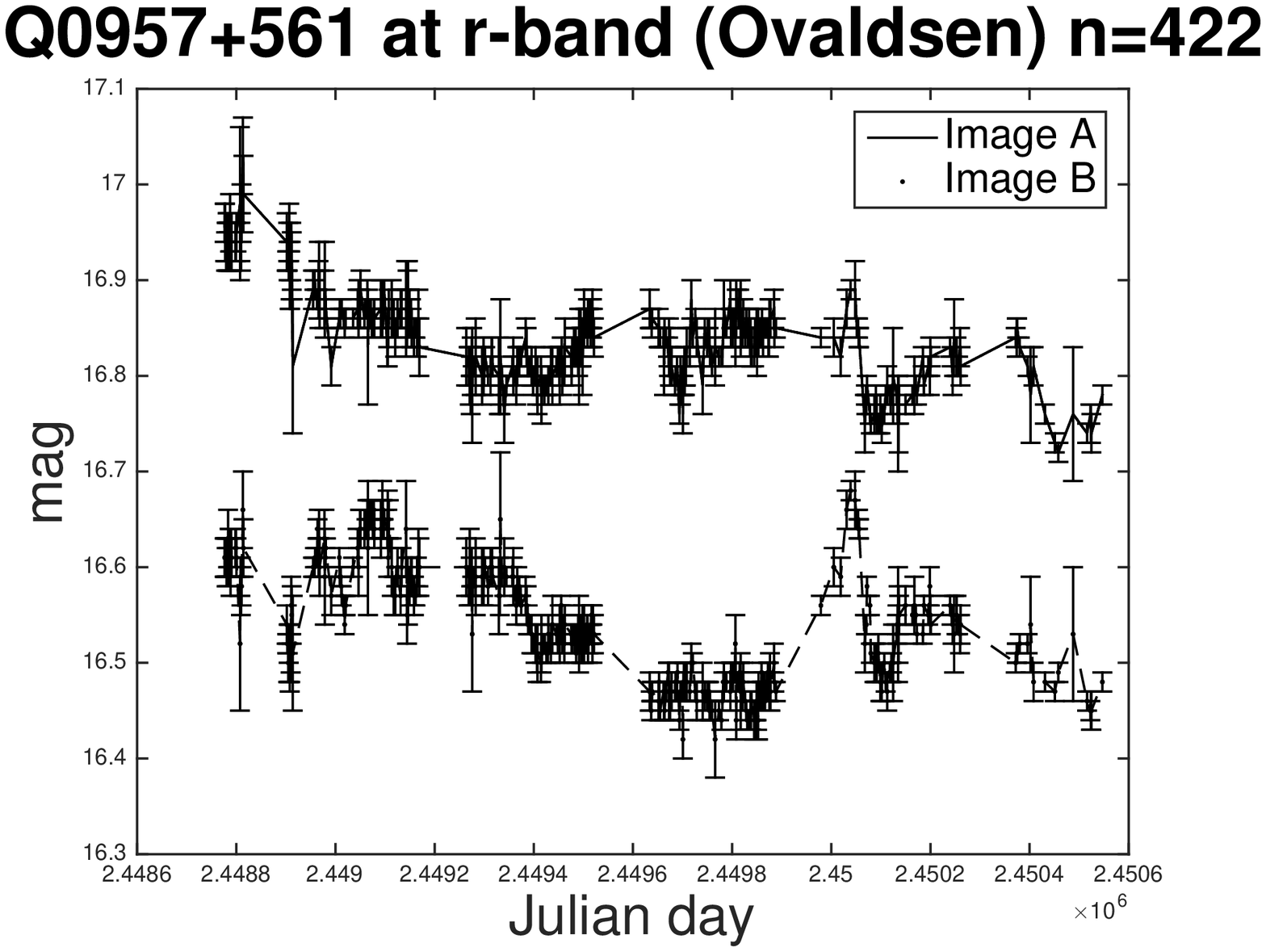}}
  \subfigure[$D^3$]{\includegraphics[width=3.1in]{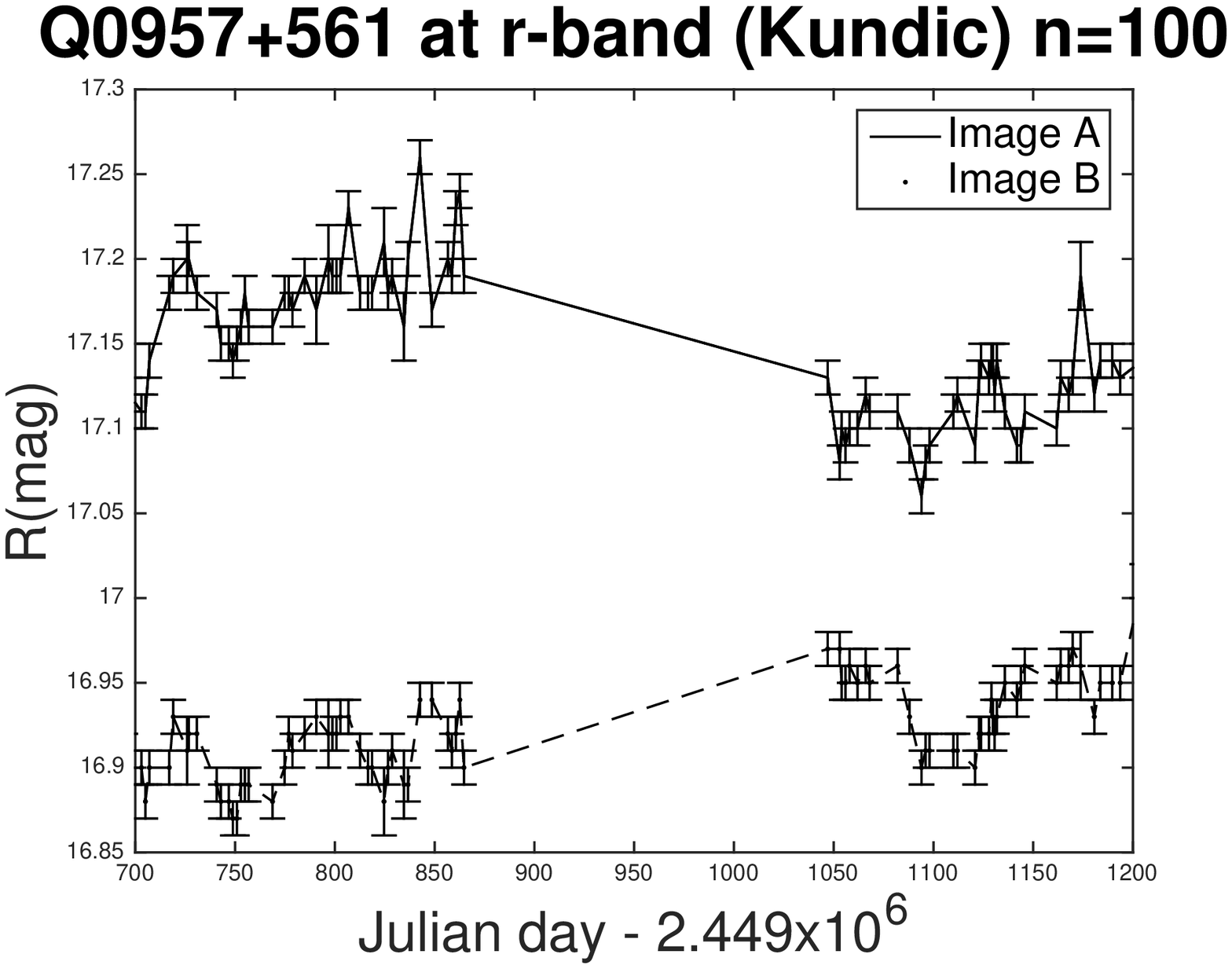}}
  \subfigure[$D^4$]{\includegraphics[width=3.1in]{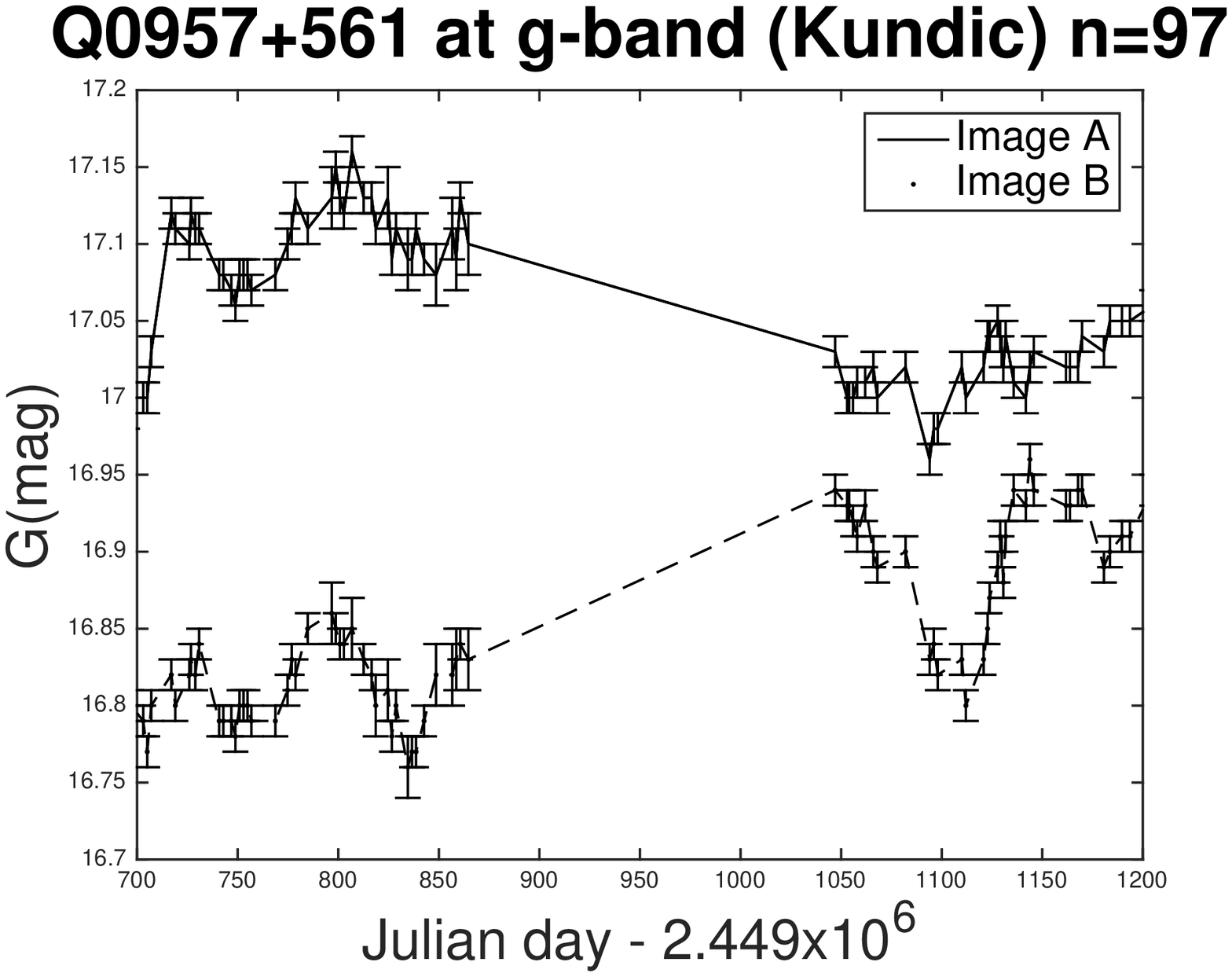}}
  \subfigure[$D^5$]{\includegraphics[width=3.1in]{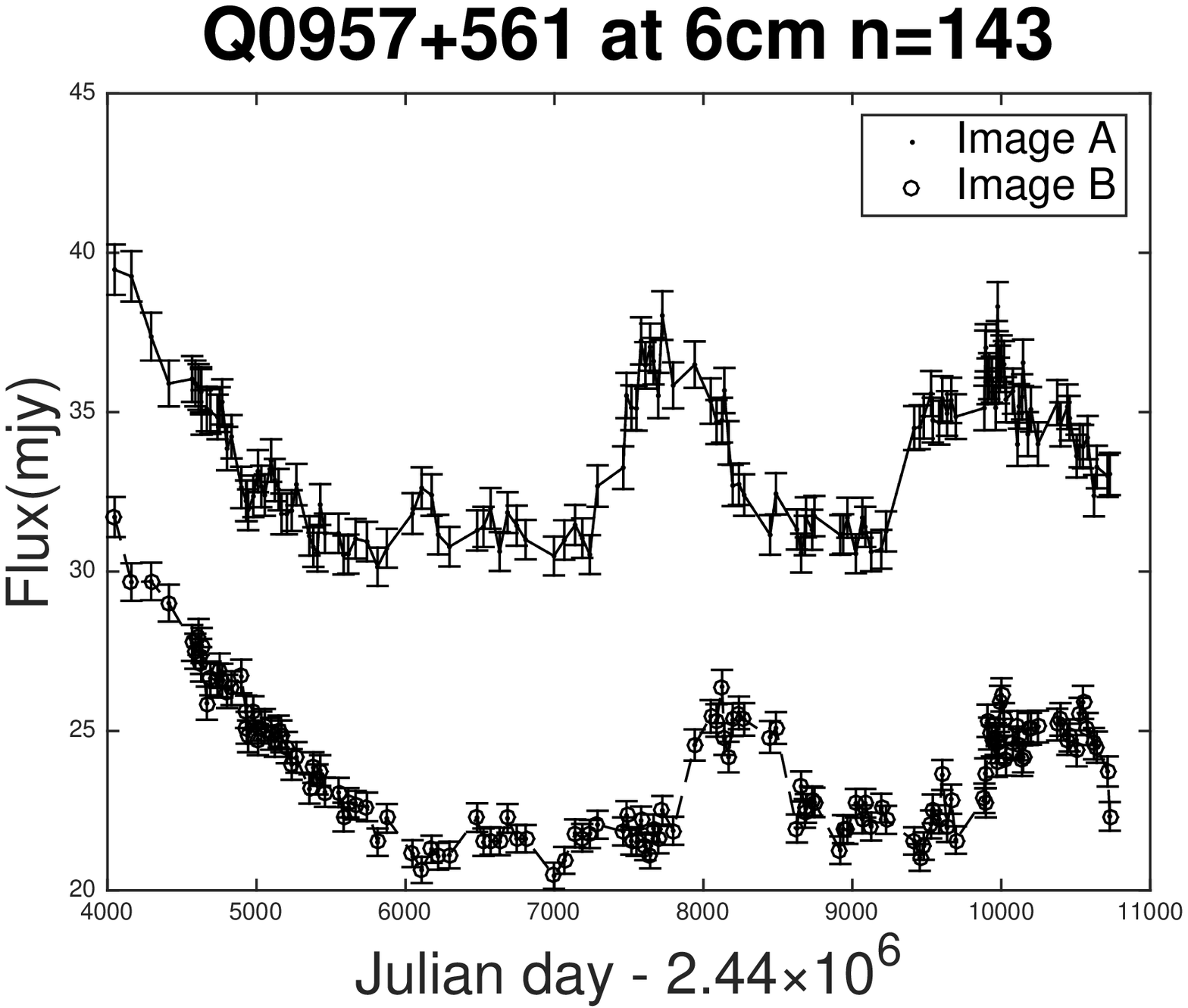}} 
  \subfigure[$D^6$]{\includegraphics[width=3.1in]{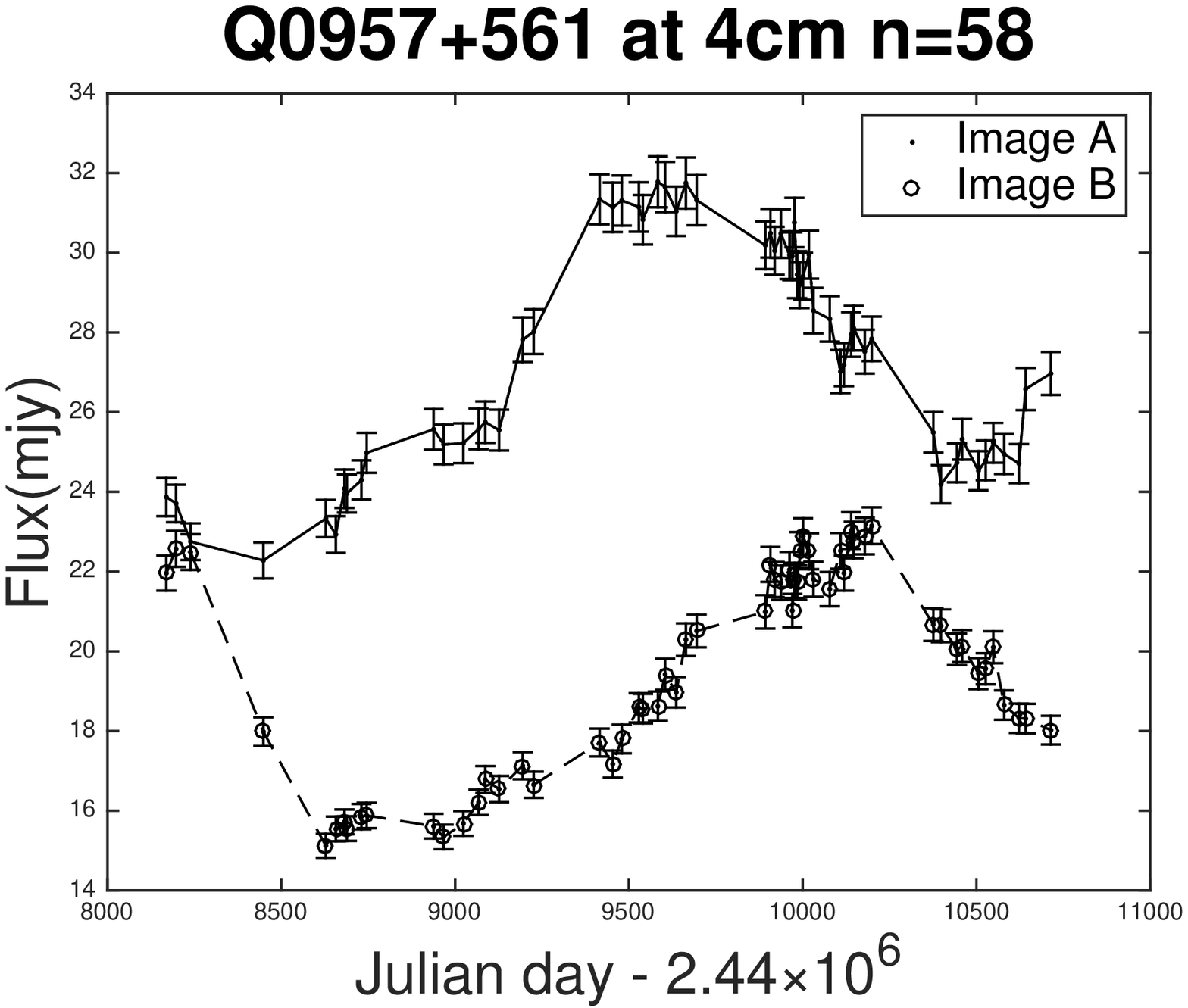}} 
 \caption{{Data set} Q0957+561. Image A from $D^1$ is shifted up by 0.6 magnitudes for clarity; image A from $D^2$ is shifted up by 0.25 magnitudes; image A from $D^4$ is shifted up by 0.05 magnitudes. For more details on these datasets see Table \ref{table:data}. }
  \label{fig:ds}
\end{figure*}

We employ six different datasets from the same quasar Q0957+561, $L=6$. The details are in Table \ref{table:data} and the plots in Figure \ref{fig:ds}.
The column labelled $N^\ell$ in Table \ref{table:data} corresponds to the number of observations per dataset. The {\it Data} column shows whether the data are optical or radio and the {\it Type} column shows the filter and the frequency used to obtain the data. The Q0957+561 is a two-image quasar, so there is either an optical magnitude offset or a flux ratio between images A and B. $D^1$ was provided by R. Schild \citep{Schild:1997:TD}, private communication.

\begin{table*}
\centering
\caption{Datasets: Q0957+561}
\label{table:data}
 \begin{tabular}{lllllll}
 \hline
   Id &       $N^\ell$  &   Data    & Type & Ratio/Offset &    Monitoring Range     & Ref\\ \hline
 $D^1$ &     1232 &    optical &  r-band   &  0.05  &  16/11/1979 -- 4/7/1998   &  \citep{Schild:1997:TD} \\
 $D^2$ &      422 &    optical &  r-band   &  0.076 &   2/6/1992 -- 8/4/1997    &  \citep{Ovaldsen:2003:NAP} \\
 $D^3$ &      100 &    optical &  r-band   &  0.21  &   3/12/1994 -- 6/7/1996   &  \citep{Kundic:1997:ARD} \\
 $D^4$ &       97 &    optical &  g-band   &  0.117 &   3/12/1994 -- 6/7/1996   &  \citep{Kundic:1997:ARD} \\
 $D^5$ &      143 &    radio   &   6cm     &  1/1.43 &  23/6/1979 -- 6-Oct-1997 &  \citep{Haarsma:1999:TRW} \\
 $D^6$ &       58 &    radio   &   4cm     &  1/1.44 &  4/10/1990 -- 22/9/1997  &  \citep{Haarsma:1999:TRW} \\
 \hline
 \end{tabular}
\end{table*}
\small
\normalsize

In order to consistently compare the performance of different time delay estimation methods in a controlled experimental setting, we also construct synthetic data on the basis of known gravitationally lensed fluxes in the optical and radio ranges, with the given observational noise and gaps structure. The ``ground truth'' - the delay - is imposed by us so that the statistics of different delay estimators can be consistently evaluated and compared.

\subsection{Synthetic Data - Realistic Experimental Setting}

\label{sec:BasicSignals}
In this section we construct synthetic signals on which we will test the proposed and some of the existing approaches to gravitational delay estimation in the presence of observational noise and gaps.
We constructed synthetic fluxes in the optical range on the basis of $D^1$ (real r-band optical data of \citep {Schild:1997:TD}) spanning roughly 10 and half years). In particular, we used $D^1$ to fit a distribution of possible fluxes `compatible' with the data (formulated as a Gaussian process) and then sampled from this distribution synthetic fluxes of 3,500 observations.

Gaussian process (GP)  represents a distribution over functions 
\begin{equation}
f(t) \sim GP (
\mu_{gp}(t),K_{gp}(t,t')),
\end{equation}
with mean and covariance functions $\mu_{gp}(t)$ and $K_{gp}(t,t')$, respectively. Any sample from the GP corresponding to a finite set of observational times $t_1,t_2,\cdots t_N$ is  Gaussian distributed  with mean
$\mu_{gp}(t_1),\mu_{gp}(t_2),\cdots \mu_{gp}(t_N)$ and covariance matrix
\small
\begin{equation}
K_{gp}=
 \begin{pmatrix}
  K_{gp}(t_1,t_1) & K_{gp}(t_1,t_2) & \cdots &K_{gp}(t_1,t_N) \\
  K_{gp}(t_2,t_1) & K_{gp}(t_2,t_2) & \cdots &K_{gp}(t_2,t_N) \\
  \vdots  & \vdots  & \ddots & \vdots  \\
  K_{gp}(t_N,t_1) & K_{gp}(t_N,t_2) & \cdots &K_{gp}(t_N,t_N)
 \end{pmatrix}.
\end{equation}
\normalsize
For our purposes, we imposed zero mean (the mean of observations in $D^1$ was shifted to zero) and used the 'squared exponential' kernel function  
\begin{equation}
K_{gp}(t, t') = \exp\left\{ - \frac{(t-t')^2}{h_{gp}^2} \right\},
\end{equation}
with scale parameter $h_{gp}$ set using cross validation on $D^1$.

A vector $(\mathbf{y},\mathbf{y}_*)^T$ of observations sampled at observation times 
$\mathbf{t}$ and $\mathbf{t}_*$ from the Gaussian process is distributed as
\small
\begin{equation}
  \begin{pmatrix}
    \mathbf{y} \\
   \mathbf{y}_* 
  \end{pmatrix}
\sim N \begin{pmatrix}
    \begin{pmatrix}
    \mathbf{0} \\
    \mathbf{0}  
    \end{pmatrix},
   \begin{pmatrix}
    K _{gp} &K_{gp*} \\
   K^T_{gp*}  &K_{gp**}
    \end{pmatrix}
  \end{pmatrix},
\end{equation}
\normalsize
where $ K _{gp}$, $ K _{gp*}$ and  $K _{gp**}$ are kernel matrices corresponding to time instances
$\mathbf{t} \times \mathbf{t}$, $\mathbf{t} \times \mathbf{t}_*$ and $\mathbf{t}_* \times \mathbf{t}_*$, respectively.
However, given observations $\mathbf{y}$ at times $\mathbf{t}$, the conditional distribution of $\mathbf{y}_*$ at times $\mathbf{t}_*$ is given by
\small
\begin{equation}
p(\mathbf{y}_*| \mathbf{t}_*,  \mathbf{y}, \mathbf{t}) = N(\mathbf{y}_*| \mathbf{\mu}_*,\Sigma_*)
\end{equation}
\normalsize
with
\small
\begin{equation}
\mu_* = K_{gp*}^T K_{gp}^{-1} \mathbf{y}
\end{equation}
\normalsize
and
\small
\begin{equation}
\Sigma_* = K_{gp**} - K_{gp*}^T K_{gp}^{-1} K_{gp*}.
\end{equation}
\normalsize


\begin{figure}
\centering
\includegraphics[width=3.2in]{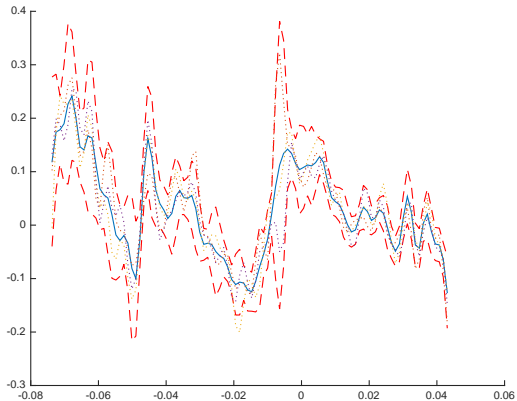}
\caption{Three Gaussian process posterior samples (dotted) based on $D^1$ (solid). Dashed curves signify $\pm$ 2 standard deviations.}
\label{fig:GPR}
\end{figure}

We sampled signals $\mathbf{y}_*$ from the Gaussian process based on $D^1$ on a regular grid of 3,500 time stamps covering the temporal range of $D^1$.  
As an example, we show three such signals in Figure \ref{fig:GPR}. Dashed curves signify $\pm$ 2 standard deviations.
To create a pair of time shifted signals A and B, the smooth long signal (signal A) $y_A = \mathbf{y}_*$ was shifted in time by a delay $\Delta=200$ days to obtain signal B, 
\begin{equation}
y_{B}(t)=y_{A}(t-\Delta).
\end{equation}
Finally, (as explained in greater detail in the following sections), we added observational noise independently to both signals A and B, and imposed observational gaps. 


%



\subsubsection{Observational Noise}
\label{sec:Noise}
Based on $D^1$ data, we first calculated the empirical distribution $p(\rho)$ of the ratio $\rho$ of the reported flux levels $y_k$ and their associated standard errors $\sigma_k$: $\rho_k = \sigma_k/y_k$.
For each observation $y(t)$ in the synthetic stream we generated an additive observational noise from a zero mean Gaussian distribution with standard deviation $\sigma(t)$, where $\sigma(t) = \rho(t) y(t)$, with $\rho(t)$ generated randomly i.i.d.~from the empirical distribution $p(\rho)$.


\subsubsection{Observational Gaps}
\label{sec:Gaps}
\begin{figure*}
 \centering
\subfigure[Objective 1]{\includegraphics[width=2.5 in]{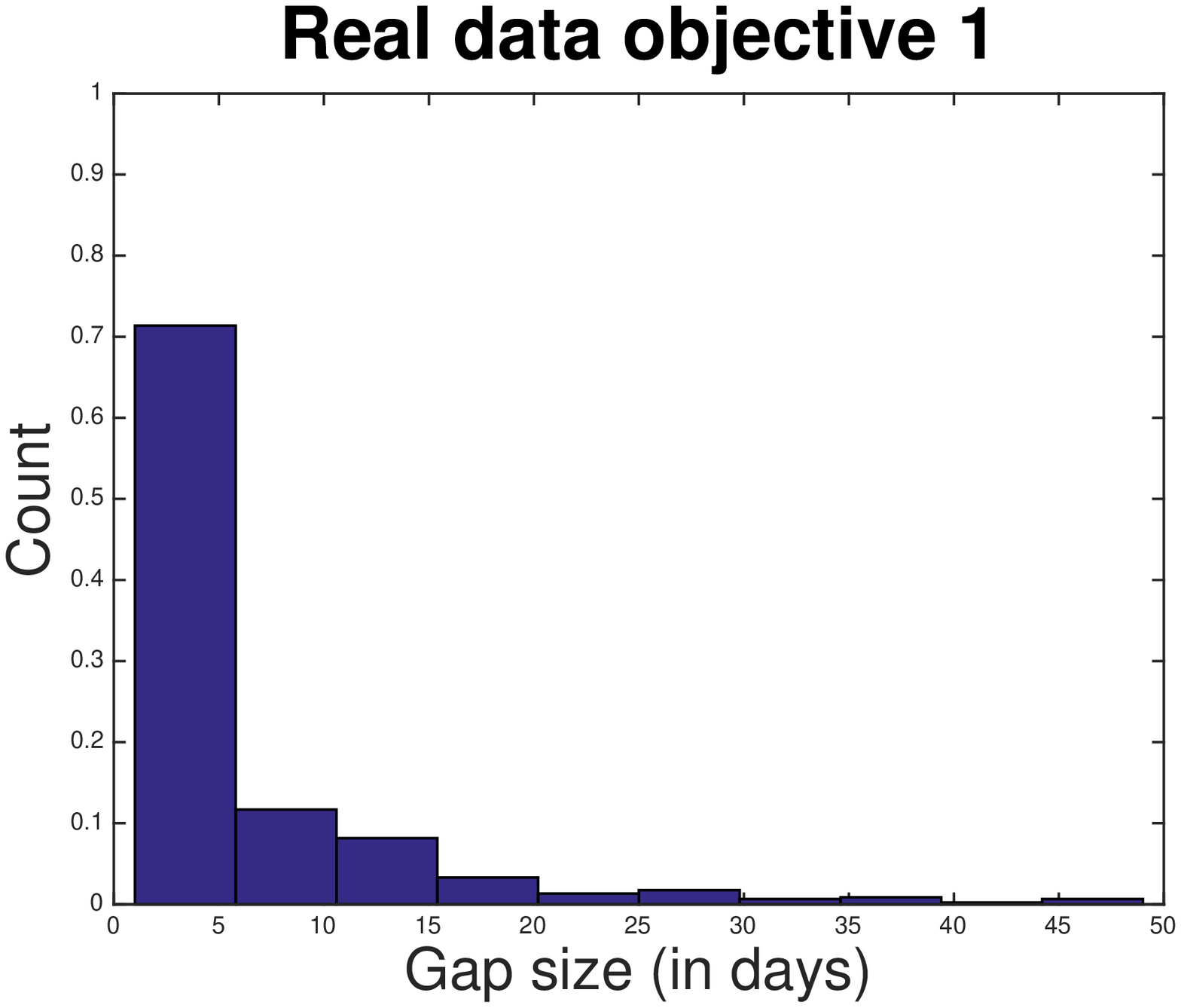}}
 \subfigure[Objective 1]{\includegraphics[width=2.5 in]{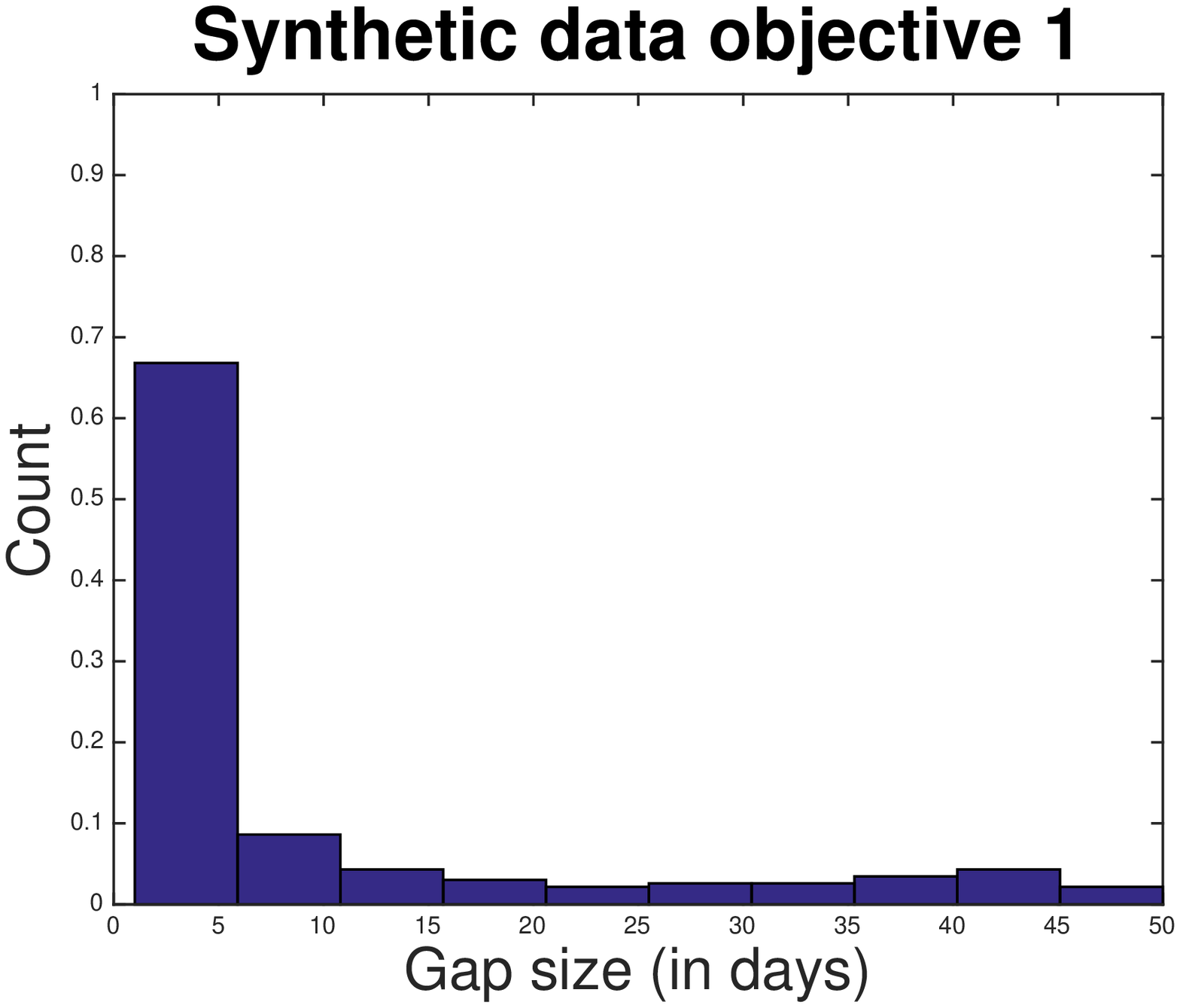}}
\subfigure[Objective 2]{\includegraphics[width=2.5 in]{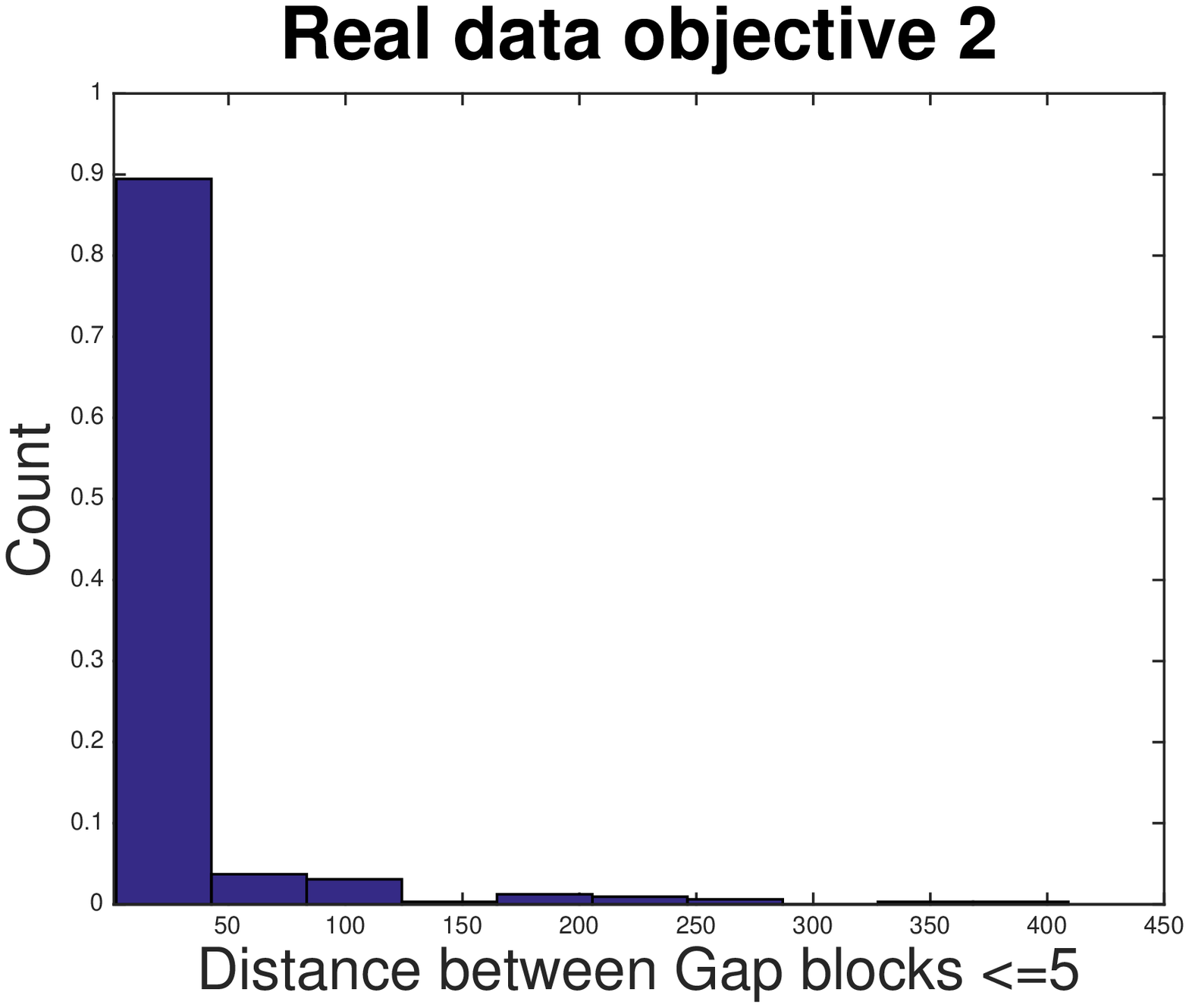}}
  \subfigure[Objective 2]{\includegraphics[width=2.5 in]{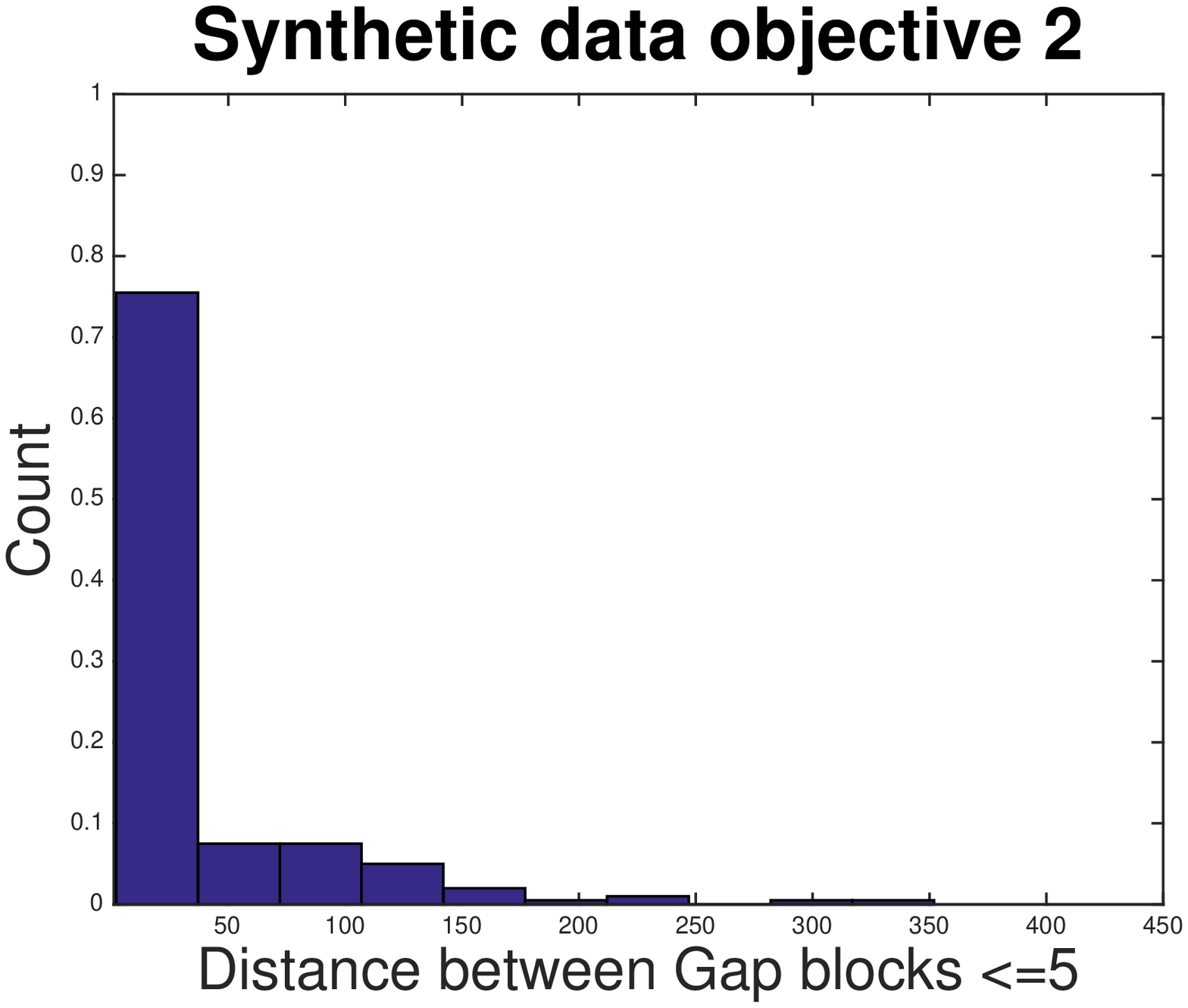}}
\subfigure[Objective 3]{\includegraphics[width=2.5 in]{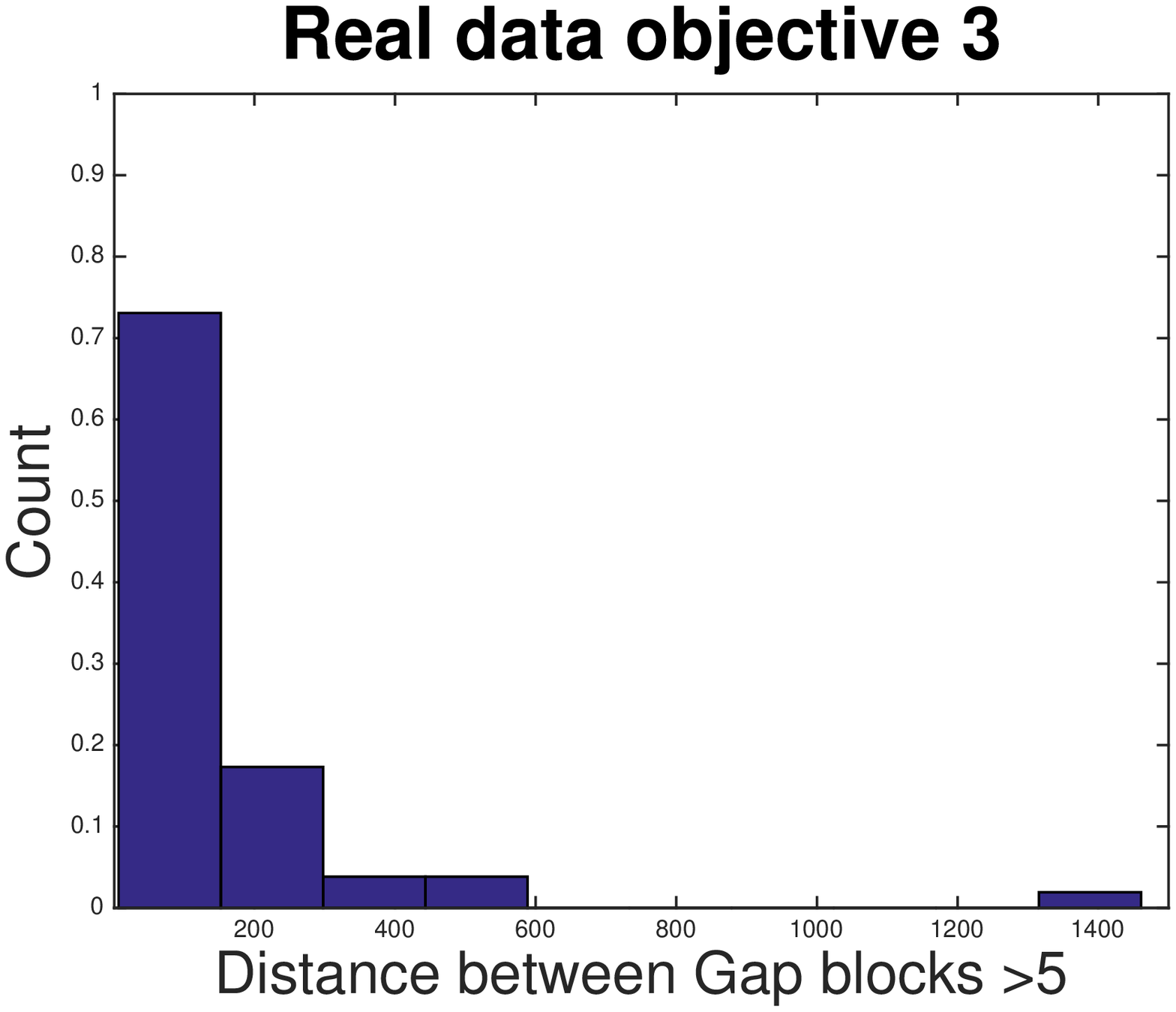}}
 \subfigure[Objective 3]{\includegraphics[width=2.5 in]{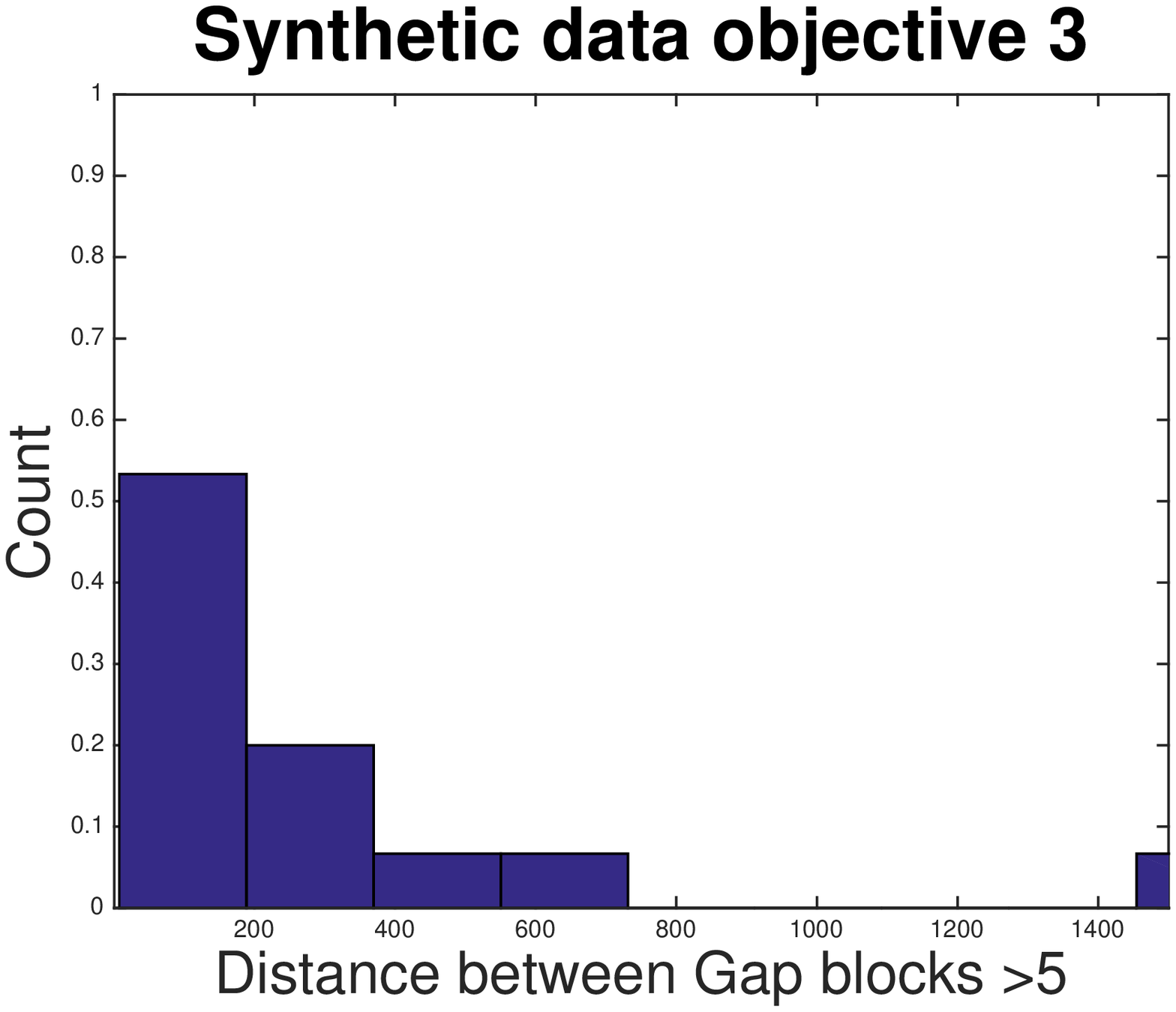}}

 \caption{Empirical distributions of gap size (a),(b),  inter-gap interval
for gaps of 1--5 days (c),(d) and inter-gap interval
for gaps of 6-10 days (e),(f). 
Each objective of RWGA corresponds to a row of two plots, left and right plots showing empirical normalized histograms from the real ($D^1$) and synthetic data, respectively. 
}
\label{fig:Ar8}
\end{figure*}

Real data are irregularly sampled due to practical considerations such as weather conditions, equipment availability, object visibility, etc.~\citep {Eigenbrod:2005:COSMOGRAIL, Cuevas-Tello:2007:tesis}. Gaps in real data are characterised by two important quantities: gap size and gap position. 
The  histogram in Figure \ref{fig:Ar8}(a) shows the empirical gap size distribution in $D^1$. Shorter gaps of 1--5 days are more frequent than longer ones (more than 6 days).

To make the synthetic data more realistic, we would like to respect constraints given by the gap size and inter-gap distance distributions for dominant gap sizes (up to 10 days). Gaps were imposed on the synthetic data by generating their sizes and positions through a multiobjective optimisation algorithm. The algorithm incorporated three constraints: (1) closeness of the generated and empirical gap size distributions; (2) closeness of the generated and empirical inter-gap interval
distributions for gaps of 1-5 days; (3) closeness of the generated and empirical inter-gap interval
distributions for gaps of 6-10 days.

The particular algorithm we used was the computationally efficient {\it Random Weighted Genetic Algorithm (RWGA)} \citep{112,111,110,113,114}.
It uses a weighted average of normalised objectives for fitness assignment (for diversity imposition the weights are randomized). The procedure is outlined in Algorithm \ref{alg:RWGA}.

\begin{algorithm}
$S=$ external archive to store non-dominated 
solutions found during
the search so far;

$n_{S}=$ number of elitist solutions immigrating from $S$ to the population of potential solutions $X_\iota$
in each generation $\iota$.

Step 1: Generate a initial random population $X_1$, set $\iota=1$.

Step 2: Assign a fitness value to each individual solution $\chi\in X_\iota$ 
by performing the following steps:

\qquad{}Step 2.1: Calculate the fitness $z_o(\chi)$ for
each objective $o=1,\ldots O$.

\qquad{}Step 2.2: Generate a random number $u_{o}$ in $[0,1]$ for
each objective $o=1,\ldots O$

\qquad{}Step 2.3: Calculate the random weight of each objective $o$
as
$w_{o}=\frac{1}{u_{o}} \sum_{i=1}^{o}u_{i}$.

\qquad{}Step 2.4: Update the overall fitness of the solution $\chi$ as $\digamma(\chi)=\sum_{o=1}^{O}w_{o}z_{o}(\chi)$.

Step 3: Calculate the selection probability $p_s(\chi)$ of each solution $\chi\in X_{\iota}$ as
follows: 
\[
p_s(\chi)=\frac{\sum_{\Upsilon\epsilon X_{\iota}}(\digamma(\Upsilon)-\digamma^{min})}
{\digamma(\chi)-\digamma^{min}},
\]
where $F^{min}=\min\left\{ \digamma(\chi)\mid \chi\in X_{\iota}\right\} $.  

Step 4: Select parents using the selection probabilities calculated
in Step 3. Mutate offspring with a predefined mutation rate. Copy
all offspring to $X_{\iota+1}$.

Step 5: Randomly remove $n_{S}$ solutions from  $X_{\iota+1}$  and add
the same number of solutions from $S$ to  $X_{\iota+1}$.

Step 6: If the stopping condition is not satisfied, set $\iota=\iota+1$
and go to Step 2. Otherwise, return to $S$.

\caption{\label{alg:RWGA}RWGA}
\end{algorithm}

The genome of each individual contains a suggestion for start positions and sizes of observational gaps. The design of individuals allows for a variable number of gaps and ensures that the gaps are not overlapping.
Figure \ref{fig:Ar8} shows the results of applying the multi-objective genetic algorithm RWGA based on $D^1$. 
Each objective corresponds to a row of two plots in Figure \ref{fig:Ar8}, left and right plots showing empirical normalized histograms from the real and synthetic data, respectively.

Generation of synthetic 'radio' data proceeded in the same way as described in the previous section for optical data, this time based on data $D^5$.

\subsection{Synthetic Data - Controlled Experimental Setting }
\label{sub:Controlled}
Generation of synthetic fluxes described above was motivated by the desire to preserve realistic gap and noise distributions. We will refer to this approach as the {\it `realistic' experimental setting} (RS). For comparing delay estimation algorithms in a large-scale controlled setting, we also considered an alternative specification of gap and noise distributions. The synthetic fluxes were first generated from the Gaussian Process model fitted to $D^1$, as described in the previous section.  The fluxes were then corrupted with observational gaps and noise. The gap sizes $g$ were generated as realisations from a mixture distribution  $P_M(g) = \alpha \cdot P_B(g; \mu_g) + (1-\alpha) \cdot P_U(g; L_g, U_g)$, where $P_B(g; \mu_g)$ is the Binomial distribution with mean $\mu_g$ and  $P_U(g; L_g, U_g)$ 
is the uniform distribution over $[L_g, U_g]$. We used the following settings: $\alpha=0.95$, $\mu_g=4,6,8$ days, $L_g=20$ and $U_g=80$.
The gap positions were randomised, subject to the constraint of minimum inter-gap distance of 2 days. The allowed range for gap size was 1 to 80 days. 
For the additive Gaussian zero mean `observational' noise we considered three settings for the standard deviation:  0.1\%, 0.2\% and 0.3\% of the flux level.
We will refer to this approach as the {\it `controlled' experimental setting} (CS).

\section{Experimental Results}\label{sec:results}
We performed experiments on synthetic datasets described in section \S\ref{sec:data}, as well as on real gravitationally lensed fluxes in the radio and optical ranges.
In the experiments we compared our methods NWE and NWE++, introduced in sections \ref{sec:model} and  \ref{sec:full_prob_model}, respectively, with two dispersion spectra approaches, namely $DS_1^2$ and $DS^2_{2,4}$ and two cross correlation approaches DCF and LNDCF.

\subsection {Experiments on Synthetic Data}
As mentioned above, we set the `true' time delay in the synthetic data to 200 days. The results of all approaches are based on testing time delay values in the range of 175 to 225 days (1 day increment). 
 
It was found that the best setting for decorrelation length $\delta$ in the $DS^2_{2,4}$ method  was 3 days.  For NWE and NWE++ the kernel width $h$ was estimated as variable kernel width with $h=2$ neighbours\footnote{
2 neighbours came consistently as the favourite option when cross-validating the number of neighbours on several initial datasets.}.
For DCF and LNDCF, the bin size is set to 5 days.
(see  \citep{Cuevas-Tello:2006:AA}). 

For each method we show the mean (bias) $\mu$ and standard deviation $\sigma$ of the maximum-likelihood delay estimates across experiments.
In all plots, the true delay is represented by the horizontal line at $\mu=200$. 

\subsubsection{Realistic Experimental Setting (RS)}
For synthetic experiments in the realistic setting we generated 500 base signals from the Gaussian process fitted to the optical data set $D^1$, as described in section \S \ref{sec:BasicSignals}. 
We then ran the RWGA algorithm to generate 500 pareto front solutions for observational gap positions and sizes (see \ref{sec:Gaps}). Each base signal thus had a corresponding observational gap structure imposed on it. Finally, the signals were corrupted by observational noise (see \ref{sec:Noise}).
The same procedure was applied for generating 500 datasets in the radio range.

\begin{figure*}
 \centering
\subfigure[Optical]{\includegraphics[width=3.1in]{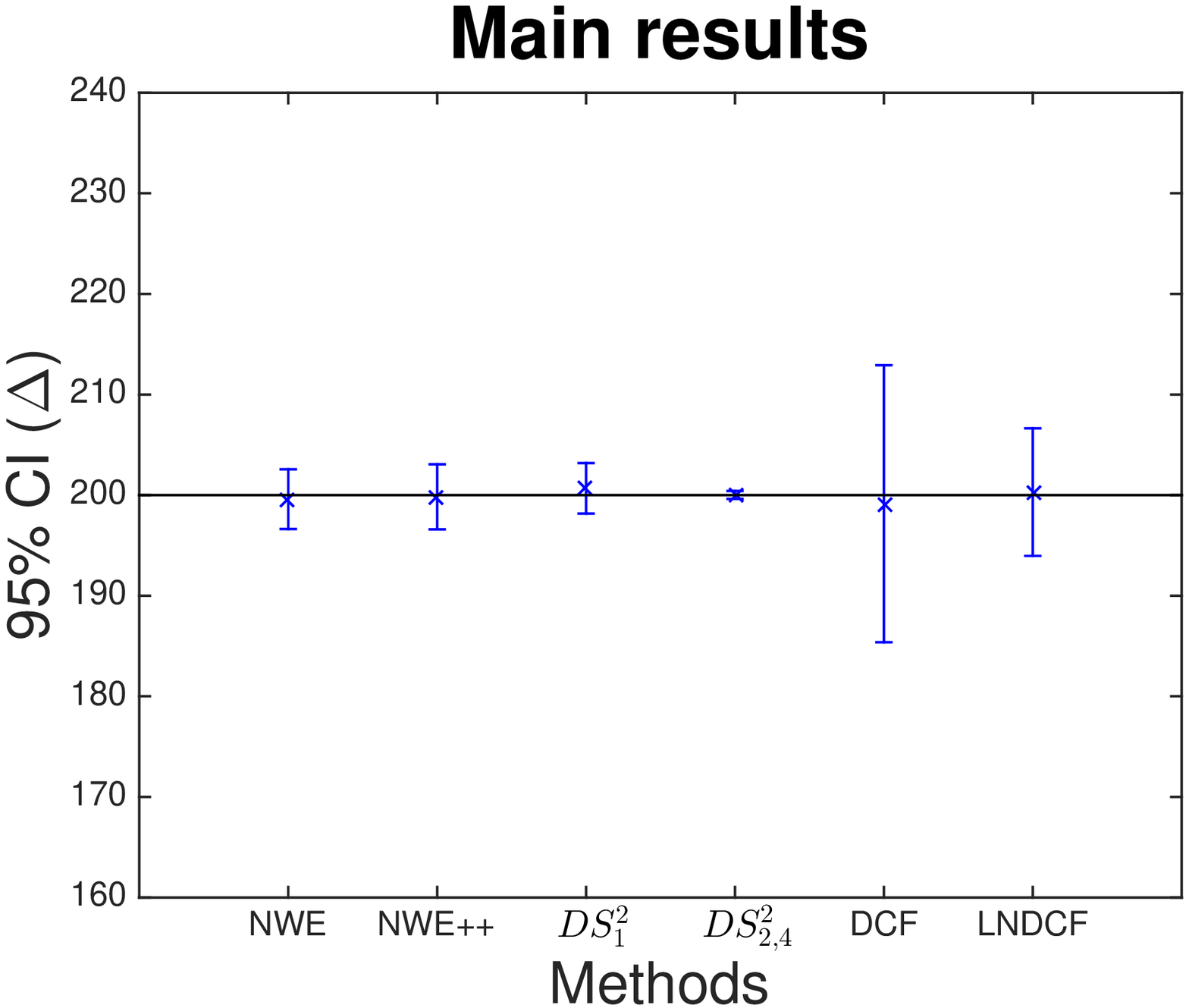}}
\subfigure[Radio]{\includegraphics[width=3.1in]{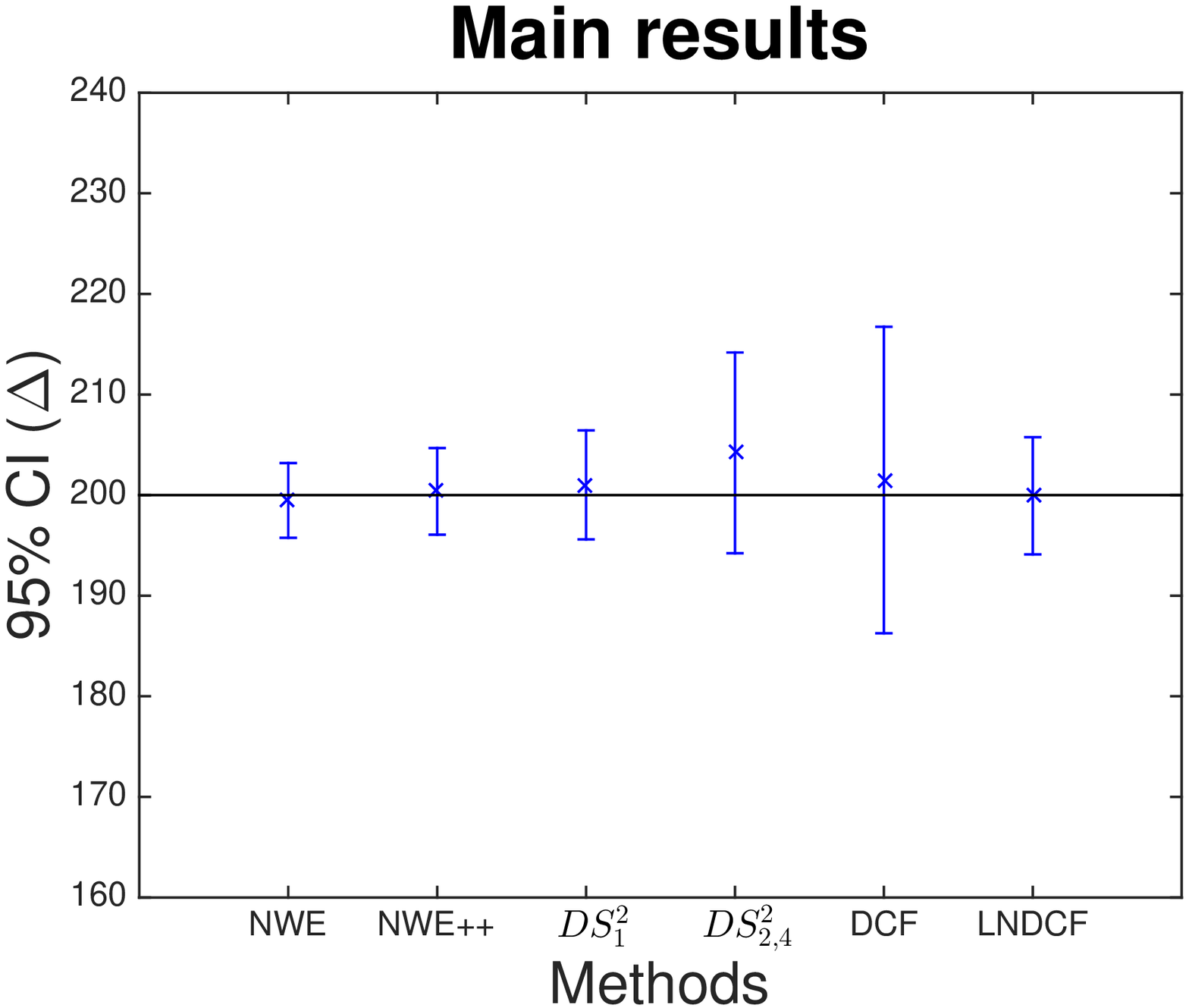}}

\caption{RS results for optical and radio data}
\label{fig:RS_EXP}
\end{figure*}

Summary results for the RS experiments on the 500 optical and radio data sets are presented in Tables \ref{tab:RS_EXP_O} and \ref{tab:RS_EXP_R}, respectively.
We report the mean ($\mu$) and standard deviation ($\sigma$) of the delay estimates $\hat \Delta_i$, $i=1,2,...,500$,
the mean absolute error (MAE) of the delay estimates (MAE$=\sum_{i=1}^{500} |\hat \Delta_i - 200|/500$), and the 95\% Credibility Interval (CI).   
The overall performance of the methods is also shown in Figure \ref{fig:RS_EXP}. 
On smaller and noisier radio data the NWE is the best performing method, followed closely by NWE++. On optical data, the best performing method is $D^2_{2,4}$.
It is important, however, to note that,in contrast to NWE methods,  the dispersion spectra methods (DS) have parameters that are difficult to set objectively based on the given data only. In the experiments, we found the best DS parameter settings by imposing the true delay $\Delta=200$, which obviously biases the DS results towards over-optimistic better performance levels. 

\begin{table*}
\centering
\caption{RS Results for optical range}
\label{tab:RS_EXP_O}
\begin{tabular}{llllll}
\hline
Method      &$\mu$\textpm $\sigma$   &MAE &CI range &95\% CI \\  
\hline
NWE         &199.60\textpm 2.97       &2.19 &0.26 &{[}199.34,199.86{]}\\
NWE++       &199.83\textpm 3.23       &2.37 &0.28 &{[}199.55,200.11{]}\\
$DS_1^2$     &200.67\textpm 2.51      &1.05 &0.22 &{[}200.45,200.89{]}\\
$DS_{2,4}^2$ &200.02\textpm 0.40      &0.16 &0.04 &{[}199.98,200.06{]}\\
DCF         &199.14\textpm 13.77      &11.61 &1.21 &{[}197.93,200.35{]}\\
LNDCF       &200.30\textpm 6.34       &4.47 &0.56 &{[}199.74,200.86{]}\\
\hline
 \end{tabular}
\end{table*}

\begin{table*}
\centering
\caption{RS Results for radio range}
\label{tab:RS_EXP_R}
\begin{tabular}{llllll}
\hline
Method      &$\mu$\textpm $\sigma$   &MAE &CI range &95\% CI \\  
\hline
NWE         &199.47\textpm 3.71       &2.95 &0.32 &{[}199.15,199.79{]}\\
NWE++       &200.37\textpm 4.31       &3.38 &0.38 &{[}199.99,200.75{]}\\
$DS_1^2$     &201.02\textpm 5.42      &4.42 &0.47 &{[}200.55,201.49{]}\\
$DS_{2,4}^2$ &204.20\textpm 9.98     &8.73 &0.87 &{[}203.33,205.07{]}\\
DCF         &201.50\textpm 15.23      &13.10 &1.33 &{[}200.17,202.83{]}\\
LNDCF       &199.94\textpm 5.83       &4.73 &0.51 &{[}199.43,200.45{]}\\
\hline
 \end{tabular}
\end{table*}


\subsubsection{Controlled Experimental Setting (CS)}
For each setting of the Binomial gap distribution $\mu_g=4,6,8$ days and for every noise level ratio from 0.1\%, 0.2\%, 0.3\%  we generated 100 base signals from the underlying Gaussian process fitted on $D^1$. We thus obtained 900 datasets. The length of the time series (after applying observational gaps) varied from 800 to 3000 observations. \par
\begin{figure*}
	\centering
	\subfigure[NWE]{\includegraphics[width=3.0 in]{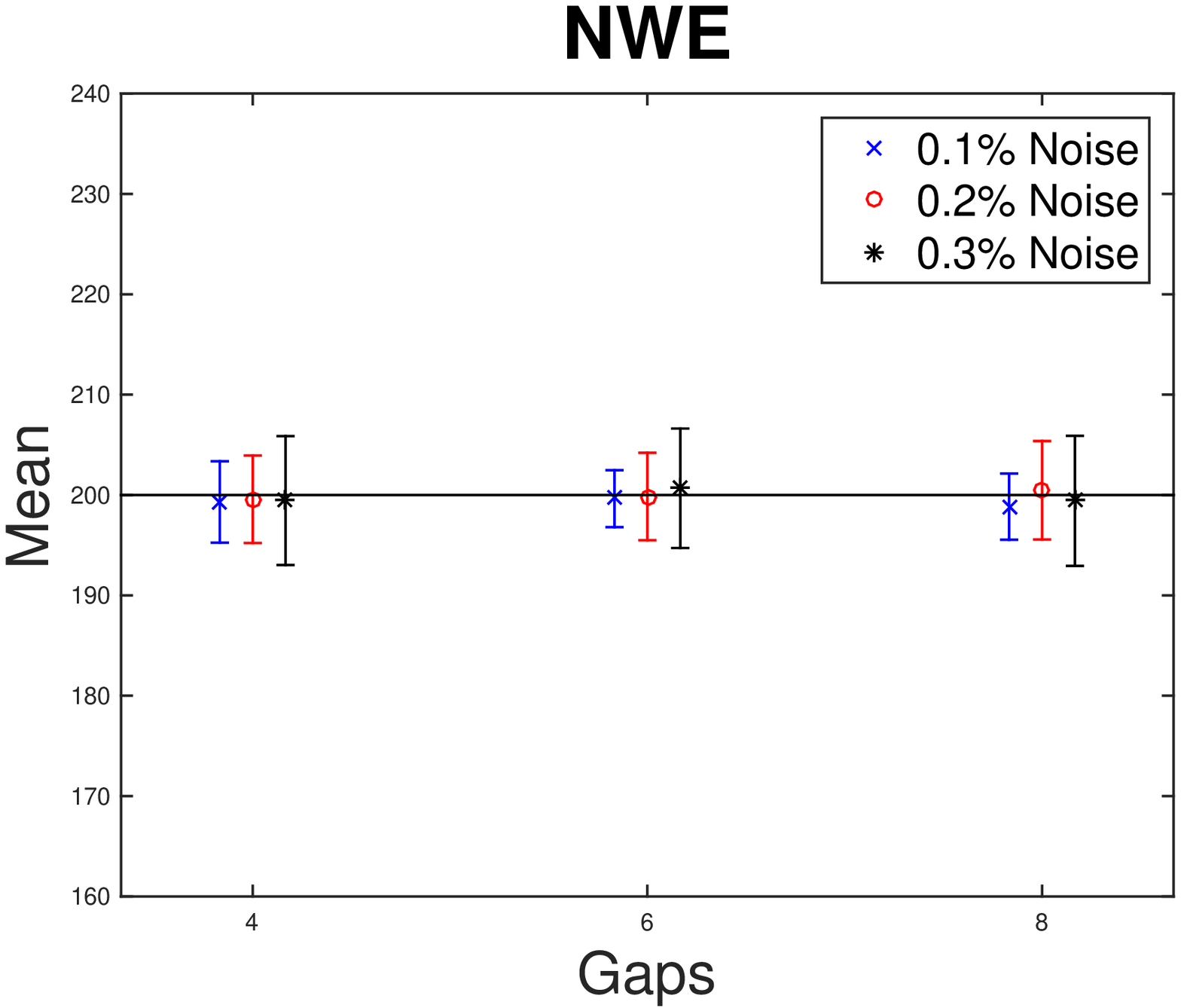}}
	\subfigure[NWE++]{\includegraphics[width=3.0 in]{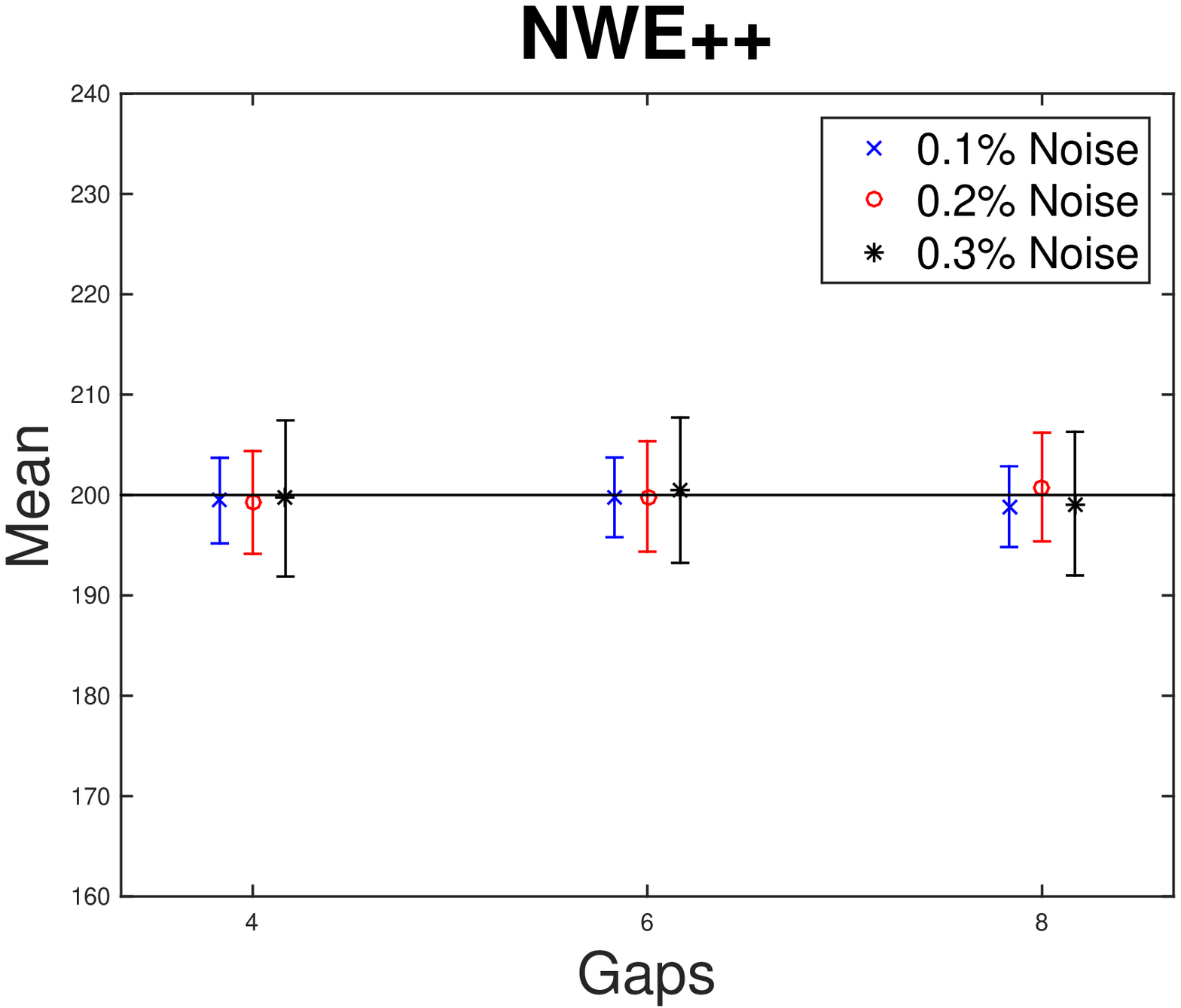}}
	\subfigure[$DS^2_1$]{\includegraphics[width=3.0 in]{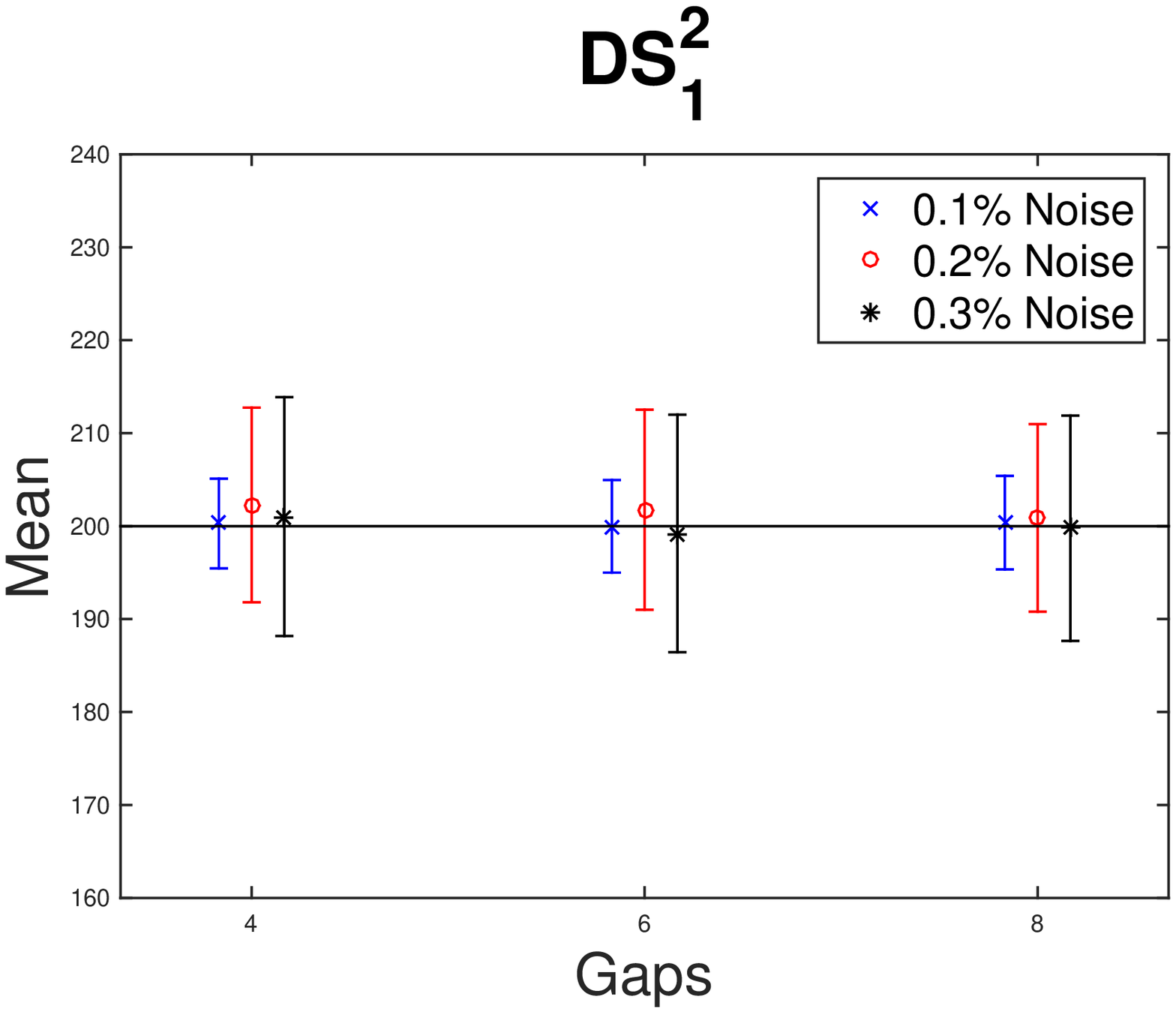}}
	\subfigure[$DS^2_{2,4}$]{\includegraphics[width=3.0 in]{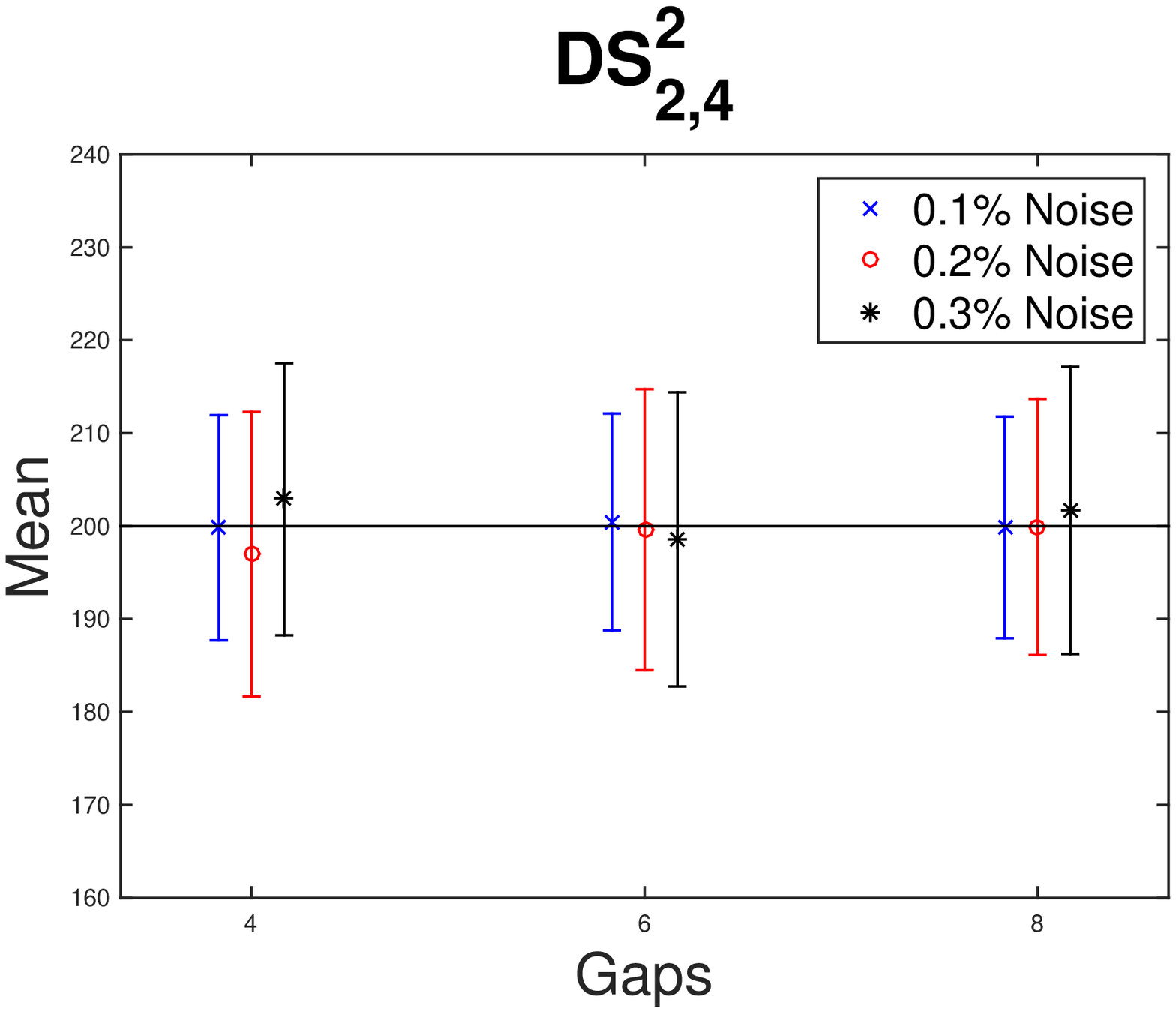}}
	\subfigure[DCF]{\includegraphics[width=3.0 in]{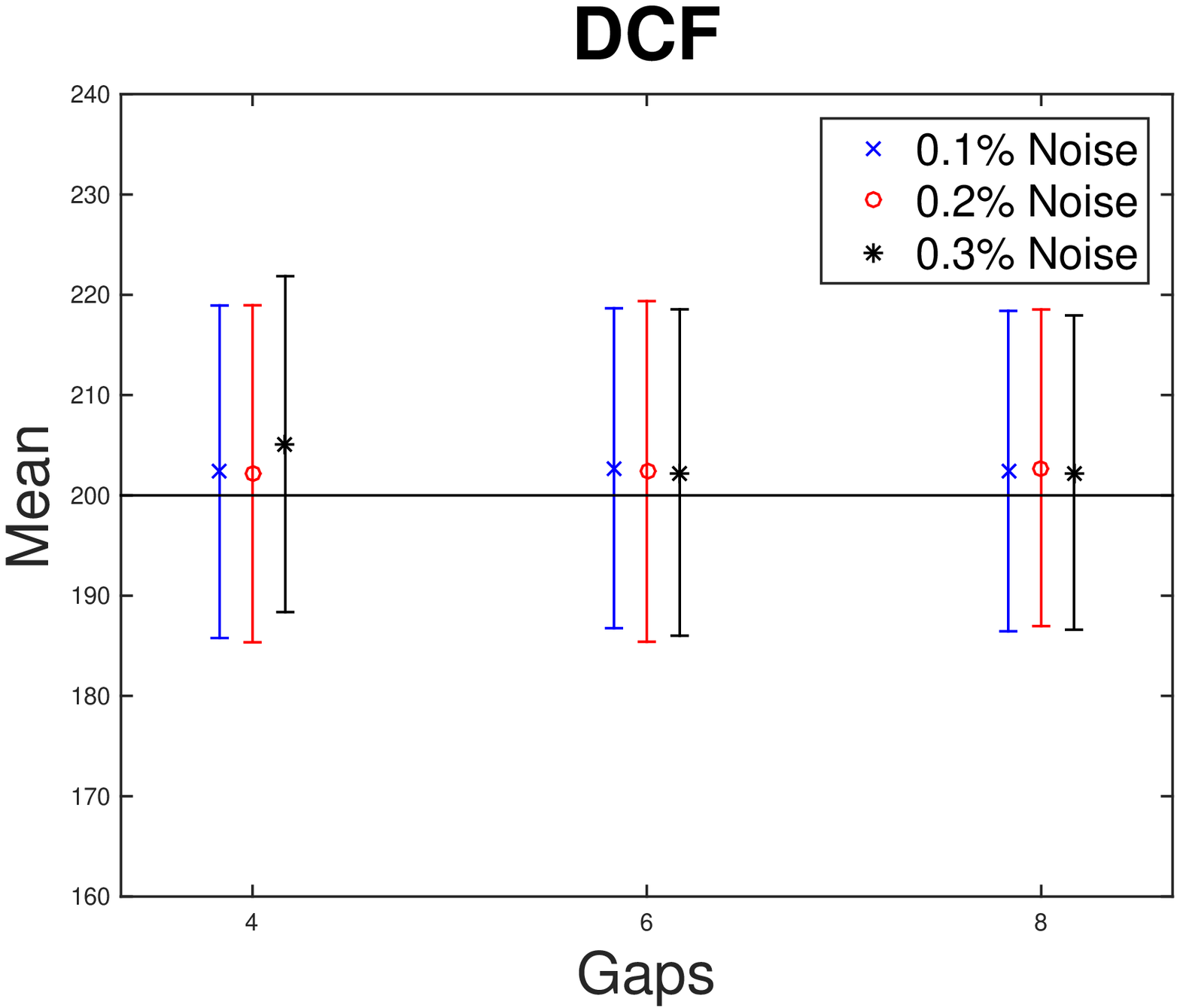}}
	\subfigure[LNDCF]{\includegraphics[width=3.0 in]{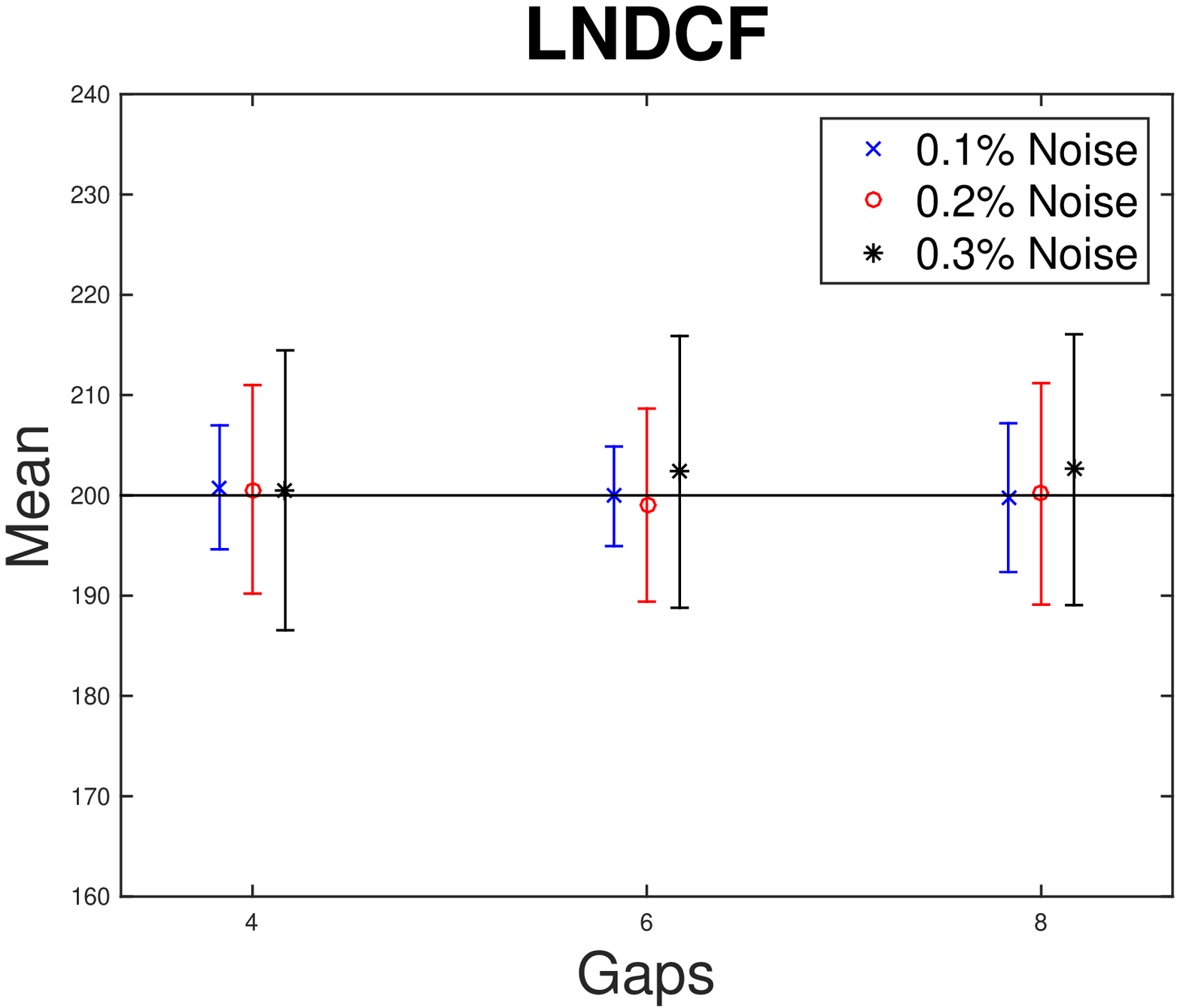}}
	\caption{CS optical range results for NWE, NWE++, $DS^2_1$, $DS2_{2,4}$, DCF and LNDCF methods (plots (a), (b), (c), (d), (e) and (f), respectively) shown as functions of $\mu_g =4,6,8$ days (mean of the binomial gap size distribution) and observational noise level. In each case we present the mean and std dev of the delay estimates for the corresponding 100 data sets.}
	\label{fig:CS_EXP_O}
\end{figure*}

\begin{figure*}
 \centering
\subfigure[NWE]{\includegraphics[width=3.0 in]{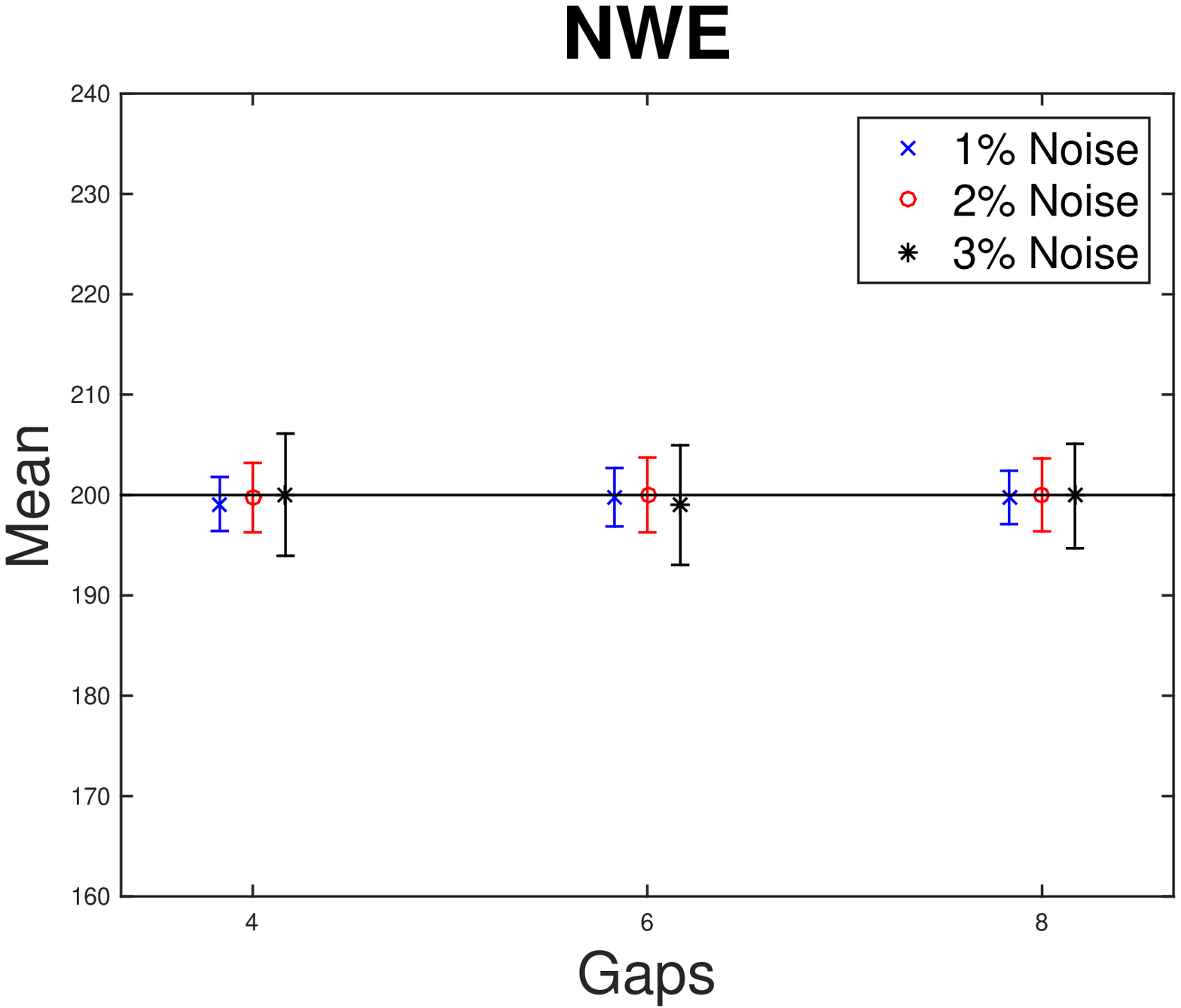}}
\subfigure[NWE++]{\includegraphics[width=3.0 in]{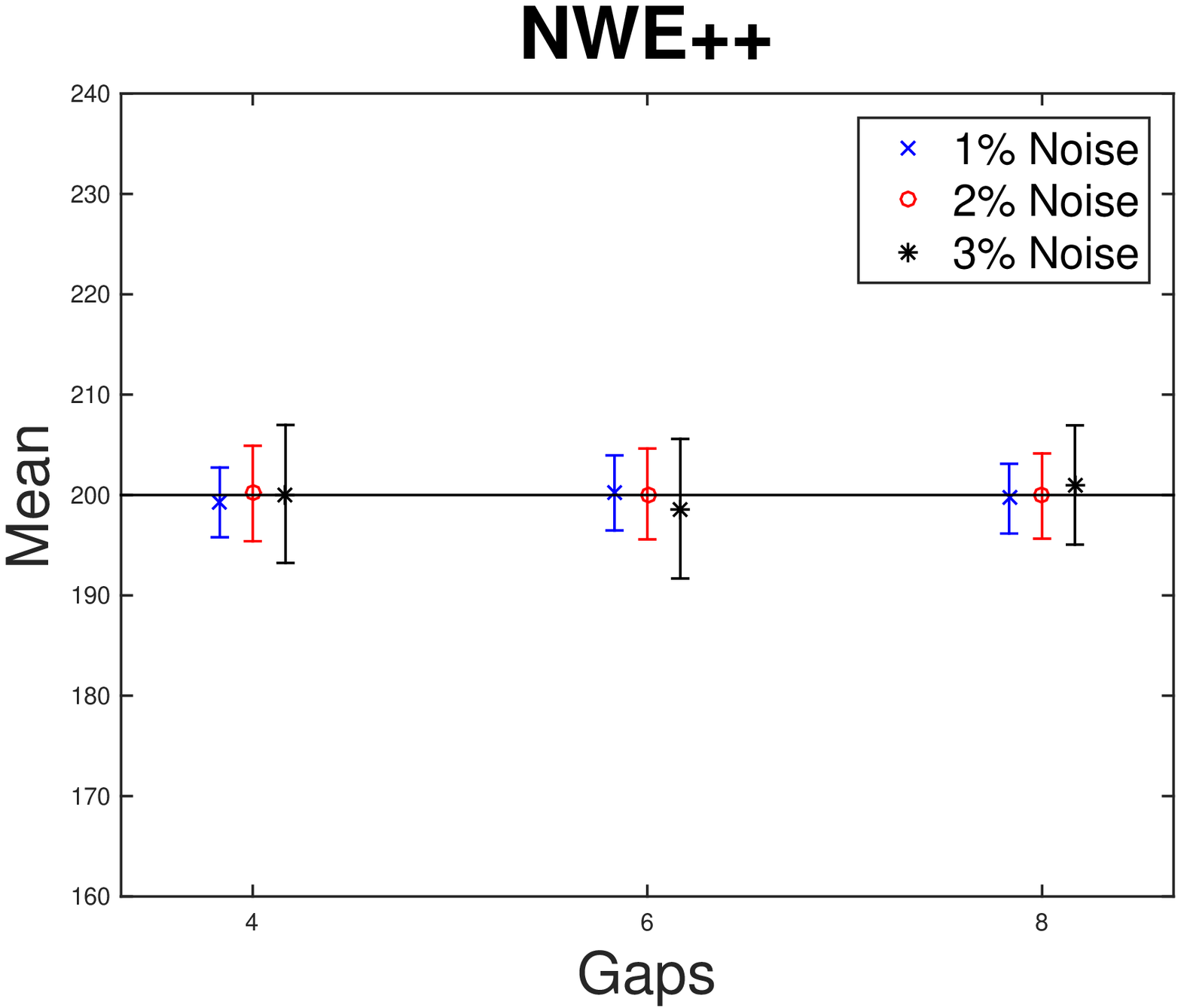}}
\subfigure[$DS^2_1$]{\includegraphics[width=3.0 in]{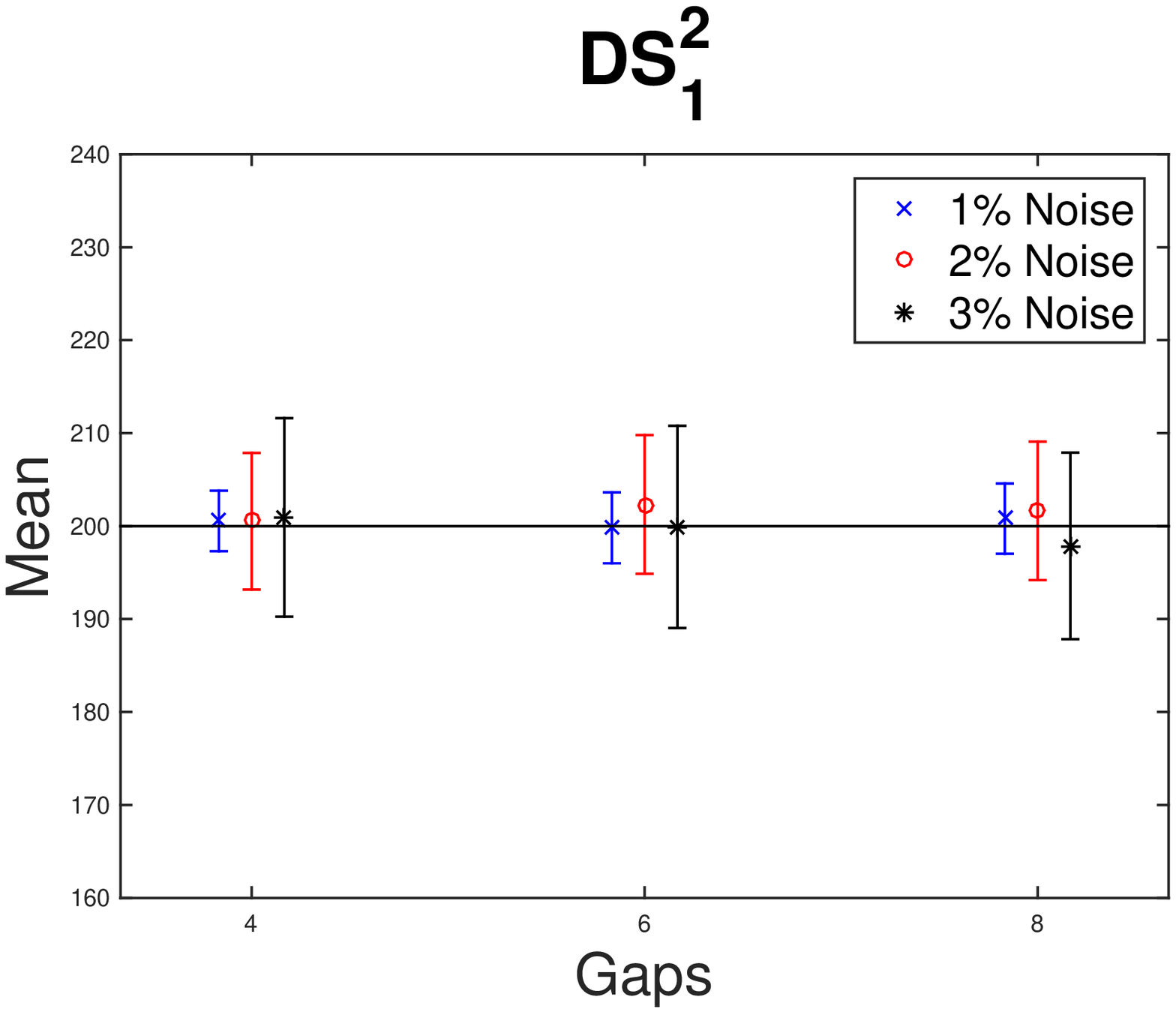}}
\subfigure[$DS^2_{2,4}$]{\includegraphics[width=3.0 in]{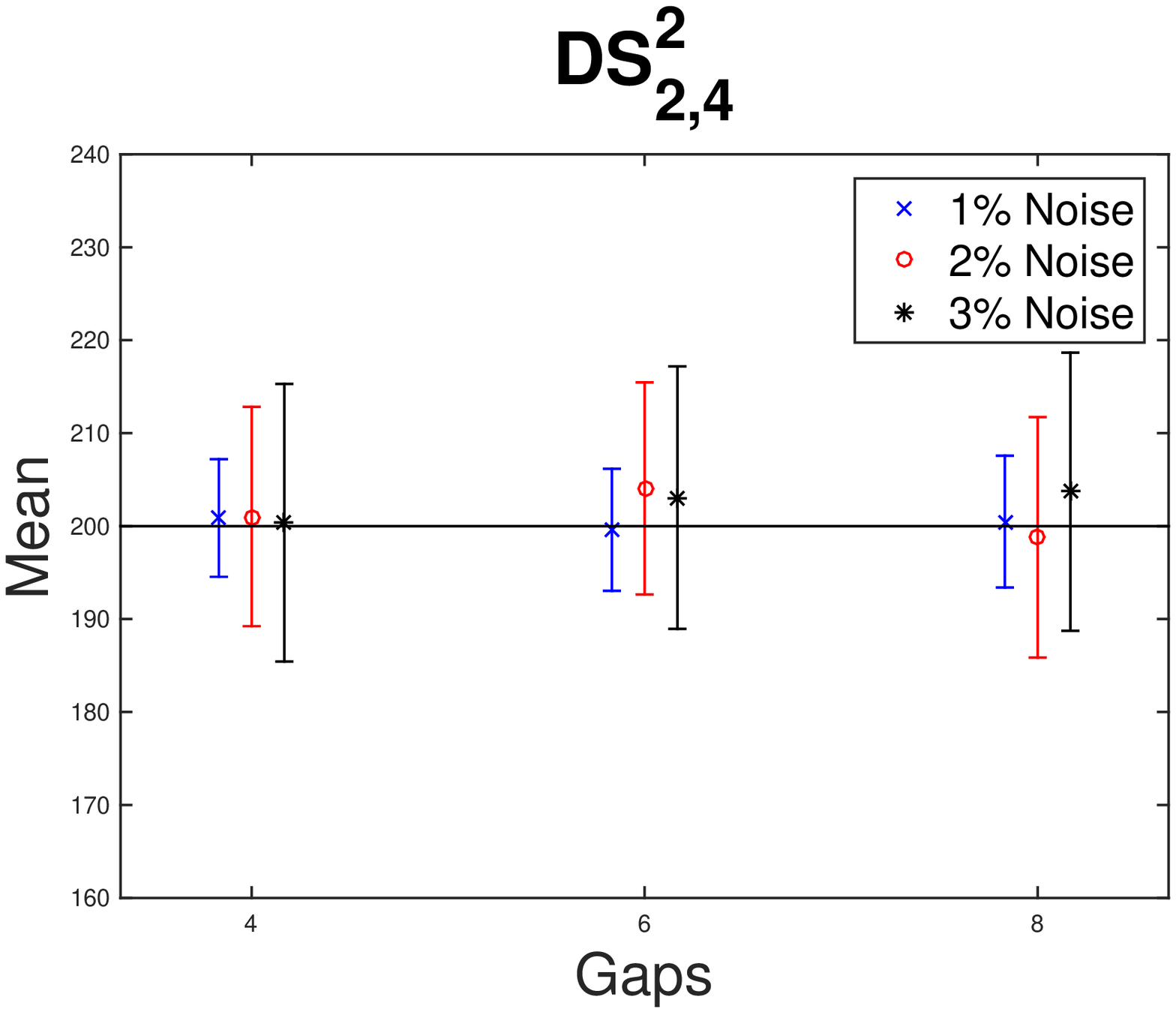}}
\subfigure[DCF]{\includegraphics[width=3.0 in]{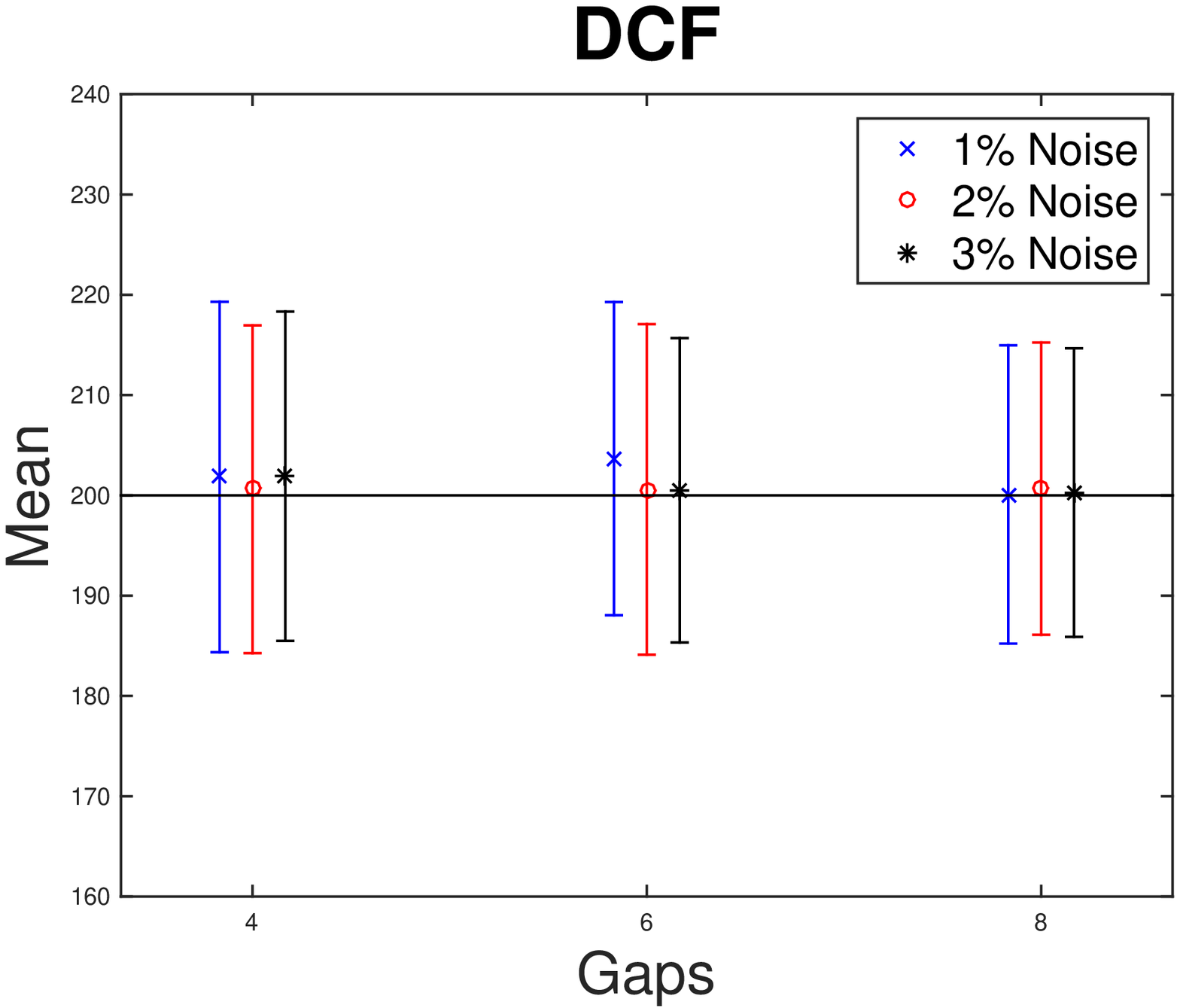}}
\subfigure[LNDCF]{\includegraphics[width=3.0 in]{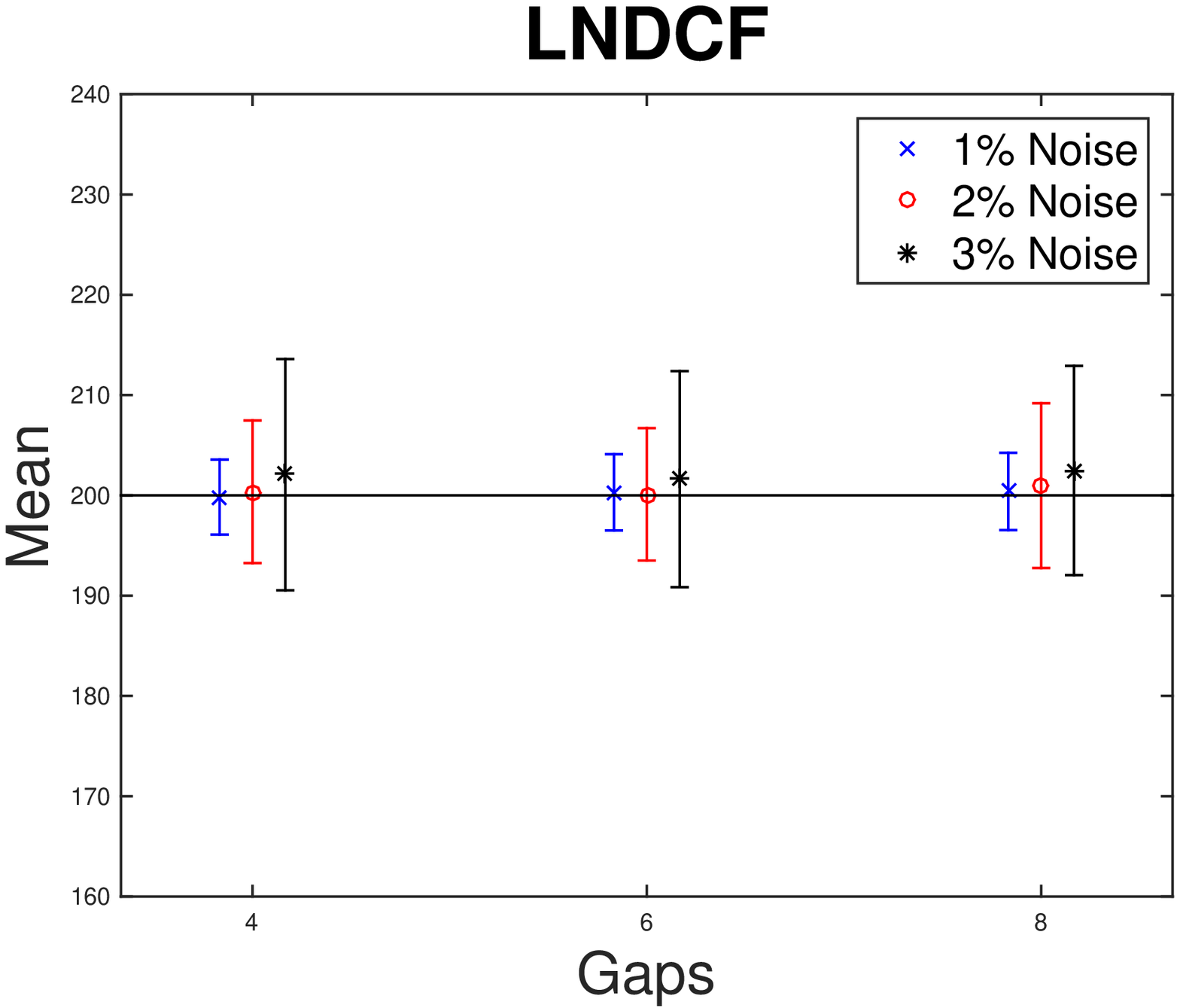}}
\caption{CS radio rang results for NWE, NWE++, $DS^2_1$, $DS2_{2,4}$, DCF and LNDCF methods (plots (a), (b), (c), (d), (e) and (f), respectively) shown as functions of $\mu_g =4,6,8 $ days  (mean of the binomial gap size distribution) and observational noise level. In each case we present the mean and std dev of the delay estimates for the corresponding 100 data sets.}
\label{fig:CS_EXP_R}
\end{figure*}

 
An analogous procedure was used to generate 900 datasets in the radio range. For each setting of the Binomial gap distribution $\mu_g=4,6,8$ days and for every noise level ratio from {1\%, 2\%, 3\% } we generated 100 base signals from the underlying Gaussian process fitted on $D^5$. 
The overall results across all CS optical and radio datasets are summarized in Tables \ref{tab:CS_EXP_O} and \ref{tab:CS_EXP_R}, respectively. 
Figures  \ref{fig:CS_EXP_O} and \ref{fig:CS_EXP_R} present the results in greater detail, grouped by noise level and gap size.

The kernel-based methods lead to more stable time delay estimates. 
NWE is the best performing method with respect to all performance measures, followed by NWE++.
It is interesting to note that while in general larger noise level ratio corresponds to larger standard deviation of the delay estimates, the DCF method seems to be more robust to increased noise levels. 
For low noise levels and with correlations between
time-shifted data streams close to unity, the DCF method is, by
construction, relatively insensitive to the level of the noise.
However, it is still clearly outperformed by other techniques for the
range of noise levels explored in this paper (see Figs. 4 and 5). 

\begin{table*}
	\centering
	\caption{Overall CS Results across all observational gap and noise settings for optical range.}
	\label{tab:CS_EXP_O}
	\begin{tabular}{llllll}
		\hline
		Method      &$\mu$\textpm $\sigma$   &MAE &CI range &95\% CI \\  
		\hline
		NWE            &199.69\textpm 4.91        &3.76 &0.32 &{[}199.37,200.01{]}\\
		NWE++       &199.69\textpm 5.78        &4.41 &0.38 &{[}199.31,200.07{]}\\
		$DS_1^2$     &200.61\textpm 9.86           &7.62 &0.64 &{[}199.97,201.25{]}\\
		$DS_{2,4}^2$ &199.97\textpm 14.10        &11.98 &0.92 &{[}199.05,200.89{]}\\
		DCF         &202.71\textpm 16.26           &14.22 &1.06 &{[}201.65,203.77{]}\\
		LNDCF       &200.63\textpm 10.56           &8.37 &0.69 &{[}199.94,201.32{]}\\
		\hline
	\end{tabular}
\end{table*}



\begin{table*}
\centering
\caption{Overall CS Results across all observational gap and noise settings for radio range.}
\label{tab:CS_EXP_R}
\begin{tabular}{llllll}
\hline
Method      &$\mu$\textpm $\sigma$   &MAE &CI range &95\% CI \\  
\hline
NWE                &199.70\textpm 4.23          &3.24 &0.28 &{[}199.42,199.98{]}\\
NWE++            &199.89\textpm 5.07          &3.90 &0.33 &{[}199.56,200.22{]}\\
$D_1^2$          &200.49\textpm 7.79          &5.92 &0.51 &{[}199.98,201.00{]}\\
$D_{4,2}^2$     &201.31\textpm 11.70        &9.36 &0.76 &{[}200.57,202.09{]}\\
DCF                 &201.13\textpm 15.70        &13.45 &1.03 &{[}200.10,202.16{]}\\
LNDCF            &200.90\textpm 7.92          &5.96 &0.52 &{[}200.38,201.42{]}\\
\hline
 \end{tabular}
\end{table*}

\subsection{Experiments on Real Data}
In this section, we present results of methods studied in this paper on real data - see Table \ref{table:data} and Figure \ref{fig:ds}. Since for real data the noise levels related to observations are available, the NWE++ method was not used.

We have $L=6$ datasets $D^1$ --- $D^6$ and for all methods, we test values for time delay on the range of $\Delta=[400,450]$ (increments of 1 day). 
The NWE cost to be minimised is $E(\textbf{h}; \Delta)$ (eq.  (\ref{eq:unique-delay})), with cross-validated  kernel scale parameters $\textbf{h} = (3, 2, 2, 2, 2, 2)$.  

For DCF and LNDCF,  the bin size $\Delta \tau$ was set to 5, 5, 5, 5, 45, and 30 for $D^1$,  $D^2$, $D^3$, $D^4$, $D^5$, and $D^6$, respectively. As mentioned before, unlike in NWE, there is no objective way of setting such parameters based on the data only and we used the setting giving most robust results in the test range of delays 400-450 days. For a fixed delay $\Delta$, the (LN)DCF function values at lag $\Delta$ are averaged across the 6 datasets $D^1-D^6$ and the combined delay estimate is obtained at the maximum of the averaged  (LN)DCF curve.

For the Dispersion spectra method $DS_{2,4}^{2}$, as argued above, the value of the decorrelation length parameter cannot be resolved in a principled manner based on the data and hence it was set to $\delta=3$, since at this value $DS_1^2$ and $DS_{2,4}^2$ have more agreement. 
Again, for a fixed delay $\Delta$, the $DS_1^2(\Delta)$  and $DS_{2,4}^{2}(\Delta)$ values are averaged across the 6 datasets  and the combined delay estimate is obtained at the minimum of such averaged  curves. The results (unique time delay across Q0957+561) are presented in Table \ref{table:Real_results}.

\begin{table}
\centering
\caption{The unique time delay across Q0957+561}
\label{table:Real_results}
 \begin{tabular}{lr}
 \hline
 Method &       $\mu$ (days)   \\ \hline
 NWE                   &420\\
$DS_1^2$           &435 \\
$DS_{2,4}^2$      &435 \\
DCF                     &408.78 \\
LNDCF                &426.31\\
 \hline
 \end{tabular}
\end{table}

To measure the uncertainty of time delay estimations, following \citep{Haarsma:1999:TRW,58,57,Ovaldsen:2003:NAP},  we also performed Monte Carlo simulations by adding white noise generated according to the reported errors to each observation\footnote{Note that this effectively adds noise to already noisy observations, resulting in a different noise distribution. For example, assuming the original noise is Gaussian, and adding random Gaussian noise from the same distribution, the standard deviation of the noise distribution in this Monte Carlo data will be $\sqrt{2}$ larger than the original one.}.
For each data set we generated 500 randomized Monte Carlo realisations. 
The results (mean and std dev across the 500 delay estimates) are presented in Table \ref{table:MC_results}.

\begin{table}
\centering
\caption{Results of 500 Monte Carlo simulations: Q0957+561}
\label{table:MC_results}
 \begin{tabular}{lrr}
 \hline
 Method &       $\mu$  (d) &   $\sigma$  (d)  \\ \hline
 NWE                  &418.65 &0.49 \\
$DS_1^2$           &434.98 &0.22 \\
$DS_{2,4}^2$      &434.92 &1.08 \\
DCF                    &408.77 &0.42 \\
LNDCF               &431.09 &15.04 \\
 \hline
 \end{tabular}
\end{table}

Although we cannot compare these results against a known true value, it is apparent that time delay estimates obtained with different methods are not mutually consistent, unlike estimates on synthetic data.  For example, $DS_1^2$ and DCF estimates appear to lie more than 50 $\sigma$ apart.  Moreover, we find that estimates using different frequency estimates on Q0957+561 data appear to be inconsistent even when the same method is used.  This suggests that the claimed measurement errors on the data are significantly under-estimated.  Alternatively, there may be unmodelled systematics (e.g., micro-lensing) that lead to varied biases for different analysis techniques.

\section{Conclusions}
\label{sec:conclusion}
We have introduced a new probabilistic efficient model-based methodology for estimating time delays between two gravitationally-lensed images of the same variable point source. The method enables one to use directly the noise levels reported for individual flux measurements.
It is more robust to observational gaps than purely  "unmodelled'' techniques, since the imposition of an identical smooth model behind multiple lensed fluxes effectively regularizes the overall model fit, and consequently, the time delay estimate itself.
 Methods such as these will be useful in the automated search for time-delay systems as well as in the accurate measurement of delays in targeted systems in future very large time-domain surveys such as those planned for the Large Synoptic Survey telescope (LSST) (e.g. \citep{Hojjati:2015,Liao:2015}).

The methods were tested and compared in two experimental settings. In the realistic setting 
the synthetic data were generated so that multiple aspects of the real data were preserved: noise-to-observed flux ratio, observational gap size distribution and the inter-gap interval distributions. The core synthetic signals were generated from a Gaussian process fitted to the real data.
In the larger controlled experimental setting the signals generated from the Gaussian process were subject to controlled levels of observational noise and gap sizes.
Our method, while being computationally efficient, showed robustness with respect to noise levels and observational gap sizes.

We also applied our method to real observed optical and radio fluxes from  quasar Q0957+561 as a combined dataset.
Of course, with real data one can estimate the variance of the estimator estimations, but never the bias, since the true time delay for Q0957+561 is not known. Our NWE estimator on the combined optical and radio data suggests a delay of approximately 420 days; however, we find that different estimators produce inconsistent results, indicating the presence of statistical or systematic measurement errors in the data in excess of the claimed measurement uncertainty.
In particular, the impact of microlensing corrections was not accounted for in the present work, and needs to be quantified in the future.

\clearpage
\section*{Acknowledgements}
The authors are grateful to Sherry Suyu, Ioana Oprea and to the anonymous reviewer for many helpful comments. Peter Ti\v{n}o was supported by the EPSRC grant EP/L000296/1. 


\bibliography{all_data}
\bibliographystyle{mn2e}
\label{lastpage}
\end{document}